\documentclass{ieeeaccess}
\pdfoutput=1 
\usepackage[utf8]{inputenc} 
\usepackage{tabularx}

\usepackage{amsmath,amssymb,amsfonts}
\usepackage{algorithmic}
\usepackage{graphicx}
\usepackage{textcomp}

\usepackage{cite}

\newcommand{\RN}[1]{%
  \textup{\uppercase\expandafter{\romannumeral#1}}%
}

\def\BibTeX{{\rm B\kern-.05em{\sc i\kern-.025em b}\kern-.08em
    T\kern-.1667em\lower.7ex\hbox{E}\kern-.125emX}}

\begin{document}

\history{Date of publication xxxx 00, 0000, date of current version xxxx 00, 0000.}
\doi{10.1109/ACCESS.2017.DOI}

\title{Modelling Metropolitan-area Ambulance Mobility under Blue Light Conditions}

\author{\uppercase{Marcus Poulton}\authorrefmark{1}, \uppercase{Anastasios Noulas}\authorrefmark{2}, 
\uppercase{David Weston}\authorrefmark{3}, and
\uppercase{George Roussos}\authorrefmark{4} \IEEEmembership{Member, IEEE}}

\address[1]{Department of Computer Science and Information Systems, Birkbeck College, University of London (e-mail: marcus@dcs.bbk.ac.uk)}
\address[2]{Center for Data Science, New York University,  (e-mail: noulas@nyu.edu)}
\address[3]{Department of Computer Science and Information Systems, Birkbeck College, University of London (e-mail: dweston@dcs.bbk.ac.uk)}
\address[4]{Department of Computer Science and Information Systems, Birkbeck College, University of London (e-mail: g.roussos@bbk.ac.uk)}

\tfootnote{The authors would like to thank Sue Money, John Downard and Leanne Smith from London Ambulance for their support and the provision of data that made this research possible.'}

\markboth
{Poulton \headeretal: Modelling Metropolitan-area Ambulance Mobility under Blue Light Conditions}
{Poulton \headeretal: Modelling Metropolitan-area Ambulance Mobility under Blue Light Conditions}

\corresp{Corresponding author: Marcus Poulton (e-mail: marcus@dcs.bbk.ac.uk).}

\begin{abstract} 
Actions taken immediately following a life-threatening personal health incident are critical for the survival of the sufferer. The timely arrival of specialist ambulance crew in particular often makes the difference between life and death. As a consequence, it is critical that emergency ambulance services achieve short response times. This objective sets a considerable challenge to ambulance services worldwide, especially in metropolitan areas where the density of incident occurrence and traffic congestion are high. Using London as a case study, in this paper we consider the advantages and limitations of data-driven methods for ambulance routing and navigation. Our long-term aim is to enable considerable improvements to their operational efficiency through the automated generation of more effective response strategies and tactics. A key ingredient of our approach is to use a large historical dataset of incidents and ambulance location traces to model route selection and arrival times. Working on the London road network graph modified to reflect the differences between emergency and civilian vehicle traffic, we develop a methodology for the precise estimation of expected ambulance speed at the individual road segment level. We demonstrate how a model that exploits this information achieves best predictive performance by implicitly capturing route-specific persistent patterns in changing traffic conditions. We then present a predictive method that achieves a high route similarity score while minimising journey duration error. This is achieved through the combination of a technique that correctly predicts routes selected by the current LAS navigation system and our best performing speed estimation model. This hybrid approach outperforms alternative mobility models. To the best of our knowledge, this study represents the first attempt to apply a data-driven methodology for route selection and the estimation of arrival times of ambulances travelling with blue lights and sirens on.
\end{abstract}


\titlepgskip=-15pt

\maketitle

\section{Introduction} 
\label{sec:intro}

The key performance indicator for emergency ambulance services worldwide is time to respond to life-threatening incidents. Indeed, a well-established clinical finding is that shorter ambulance arrival times are critical for the survival of individuals involved in high-severity emergency incidents \cite{valenzuela_estimating_1997}. In the United Kingdom for example, ambulance services operating under the National Health Service (NHS) treat approximately $30,000$ out-of-hospital cardiac arrest (OHCA) patients every year \cite{perkins_uk_2015} with a survival rate of approximately $9\%$ for those reaching hospital discharge \cite{daya_out--hospital_2015}. Short ambulance response times are a key factor in improved clinical outcomes because longer delays between collapse and the commencement of emergency life support result into significantly reduced survival rates \cite{weisfeldt_resuscitation_2002}. After $10$ minutes very few patients survive \cite{graham_emergency_2015}.

Accordingly, emergency ambulance services in the UK operate under the regulatory requirement to reach at least $75\%$ of OHCA patients within $8$ minutes. Yet, in major metropolitan areas such as London meeting this mandate presents considerable challenges, especially in a setting of horizontal cuts in funding and a rising number of medical emergencies: in October 2017, only $68\%$ of patients were treated within the time frame required. 

To address the challenges of effective ambulance response, in this paper we introduce a data-driven methodology with a view to enable considerable improvements to the operational efficiency of  emergency services. Specifically, we investigate the routing and navigation performance of a metropolitan ambulance response system using London as our case study. Our main goal is to exploit the improved understanding of mobility patterns as relating specifically to ambulances, so that we can model and predict their movements precisely. Our longer-term goal is to contribute towards the generation of more effective response strategies and tactics through software automation. 

We investigate the performance of this data-driven approach using a large historical dataset collected specifically for this study from the London Ambulance Service (LAS) over a $2$-year period. This dataset was assembled by the lead author with full access to LAS telemetry (described in more detail in Section \ref{section:background}) which captures ambulance movements in the city as the service responds to emergency calls. The dataset offers a distinct view of a metropolitan-scale emergency response system at a level of detail and at scale not previously considered in the research literature. To the best of our knowledge, this study introduces the first data-driven methodology for algorithmic route selection and the estimation of travel time specifically tailored to emergency ambulatory vehicles responding to an incident with blue lights and sirens on.

In Section~\ref{sec:ambulancemobility}, we describe the incident and ambulance tracking dataset and then proceed to employ ambulance journey data to enrich emergency vehicle-specific graph representations of London's road network reflecting systematic spatio-temporal variations that characterise the speed patterns of emergency vehicles. We show how these speed variations capture implicit, yet essential, information, of underlying traffic conditions. Taking these into account can effectively lead to better estimates of ambulance arrival times at the location of an incident, which is key in predictive modelling and dispatch simulation. 

We summarise our research findings with the following key points:

\begin{itemize}

\item \textbf{Spatio-temporal analysis of ambulance mobility:} 
Our analysis of ambulance movement patterns in Section~\ref{sec:ambulancemobility} reveal that, in line with London's well known polycentric urban structure, demand for the service concentrates around several urban nuclei within the metropolitan area of the city. Moreover, we identify strong temporal variations in the speed of different types of ambulatory transport vehicles. Finally, we demonstrate that ambulance mobility significantly differs from civilian traffic and thus requires a specifically tailored approach. 

\item \textbf{Multi-layer graph representation of London's road network for Blue Lights traffic:} Reflecting the fact that emergency vehicles that travel with blue lights and sirens on are exempt from traffic regulations, we curate a graph representation of London's road network in a manner that accommodates the distinctive requirements of emergency services. The Blue Lights Road Network described in detail in Section \ref{section:roadnetwork} permits right turns where restrictions apply on civilian traffic, among other characteristic features. Furthermore, each road segment represented in the network is associated with a set of weights employed to provide alternative ambulance speed estimates.  

\item \textbf{Accurate ambulance speed estimation from low-frequency GPS data:}
A key element of the data-driven approach presented in this paper is the calculation of accurate expected ambulance speeds on every segment of the London road network. To achieve this, in Section~\ref{section:map-matching} we employ the Blue Lights Road Network and a map-matching method selected for its performance on coarse-grained GPS ambulance tracking data to reconstruct complete ambulance routes. Fully reconstructed routes are subsequently employed to obtain considerably improved speed estimates on a per road-segment basis thus enabling the calculation of more precise expected ambulance journey times.

\item \textbf{Development of a data-driven predictive ambulance mobility model:}
Building on the above findings, we develop and evaluate alternative predictive ambulance mobility methods of increasing sophistication. First, we consider journey time accuracy as the key success criterion, concluding that the best performing model is adaptive to both time of week and the specific set of road segments traversed (Section~\ref{section:modelling}). We demonstrate through in-depth route similarity analysis, that a hybrid model combining a route selection stage followed by travel time estimation can improve intelligence regarding journey times, specifically capturing the design features of the navigation system currently in place at the service (Section~\ref{section:hybrid}). Last but not least, we identify opportunities for improved selection of ambulance routes.
\end{itemize}

\section{Background and Rationale} 
\label{section:background}

Setting the context for the detailed description of the dataset used in this study, in this section we provide a concise operational overview of the LAS, including a description of the mobile computing technology implemented on-board ambulances. We then proceed to demonstrate the features that differentiate emergency vehicle movement and general civilian traffic and conclude with a brief overview of relevant literature.

\subsection{LAS Overview}
The LAS is responsible for providing emergency medical care across Greater London, covering an area of approximately $1,572\:\mathrm{km\textsuperscript{2}}$ with a population of 8.77 million in 2016. Indeed, London has the highest population density in the United Kingdom with $5,235$ residents per square kilometre. In 2015, LAS attended $10,211$ out-of-hospital cardiac arrest (OHCA) incidents and attempted to resuscitate $4,665$ of the patients involved. During 2010-15, LAS experienced a persistent increase in the volume of high-priority calls, which accounted for $40\%$ of all recorded incidents. 

LAS employ several vehicle types within their fleet consisting of approximately $400$ ambulance units (AEU), $200$ fast response units (FRU) and a smaller number of bicycles (CRU) and motorcycles (MRU). AEUs in particular are typically commercial trucks modified to provide facilities for emergency treatment en route to an appropriate Accident \& Emergency ward. 

Ambulance crew can have different levels of skill and experience. Emergency Ambulance Crew (EAC) and Emergency Medical Technicians (EMT) are trained to respond quickly to out-of-hospital medical emergencies. They are able to deliver a wide range of treatments to patients such as those suffering from cardiac arrest, trauma or minor injuries, and are authorised to deliver drugs and provide immediate life support such as defibrillation and basic airway management. Paramedics are authorised to conduct invasive procedures such as advanced airway management and needle chest decompressions, as well as deliver a wider range of drugs. Ambulance crews typically work on $12$-hour shifts beginning at 7am and 7pm respectively. Each crew can take a single 30-45 minute rest break during a ``rest break window.''

Since April 2011, the Department of Health (DoH) have set targets for ambulance trusts such that $75\%$ of immediately life-threatening incidents must have a first responder arrive within $8$ minutes and $95\%$ of patients must be reached within $19$ minutes \cite{hsr_department_at_scharr_university_of_sheffield_comparative_2009}. 
Note that these targets are relatively lower than international standards, which typically require sub-$8$ minute response for $90\%$ of immediately life-threatening cases. Indeed, the OPALS study calls for a maximum $5$ minute arrival time to $90\%$ of Category A cases \cite{stiell_advanced_2004}. Other non-life threatening incidents have locally agreed targets typically requiring the first responder to arrive within $20$ to $30$ minutes depending on the severity. Table \ref{tab:targets} summarises target response times. 

\begin{table}[t]
	\centering
	\caption{UK DoH ambulance response targets.}
	\label{tab:targets}
	\begin{tabular}{|p{28pt}|p{85pt}|p{75pt}|}
	\hline
	{\bf Category} & {\bf Response time} & {\bf Comment}  \\ \hline
	\emph{Red 1} & Within 8 min  & Cardiac arrest \\ \hline
	\emph{Red 2 }& Within  8 min  & Other life threatening emergencies  \\ \hline
	\emph{Green 1} & Within 20 min  & Blue lights and sirens   \\ \hline
	\emph{Green 2} & Within 30 min  & Blue lights and sirens   \\ \hline
	\emph{Green 3} & Assessment within 20 min  & Response within 1 hour\\ \hline
	\emph{Green 4} & Assessment within 1 hour  & \\ \hline
	\end{tabular}
\end{table}

\subsection{On-board Technology}
AEUs and FRUs are equipped with an on-board computer known as the Mobile Data Terminal (MDT) incorporating a touch-sensitive display. Information transmitted to the MDT by the LAS Command and Control Centre is used to provide the crew with patient and incident details, and map-based search and navigation facilities. AEUs also carry extensive instrumentation that tracks their location via GPS, and monitors vehicle state including equipment temperature, handbrake position, door open, blue lights, sirens, batteries, fuel level and so forth. This information is relayed periodically to LAS headquarters over multiple wireless pathways including at least two 2G cellular mobile telephony networks to ensure resilience and extended coverage, as well as IEEE 802.11 when an ambulance is near an ambulance station. Similar to other UK emergency services, ambulances also carry two-way TETRA transceivers which carry encrypted voice and data so that the crew can communicate directly with the LAS Command and Control centre. 

As relates to location information specifically, AEUs  carry a Siemens Navigation Unit (SNU) which provides geographic positioning information and routing capabilities to the MDT. The SNU incorporates embedded GPS receivers, gyroscopes and accelerometers and receives supplementary information from external wheel speed sensors. Positioning information is recorded whenever the vehicle is en-route to an incident, A\&E, fuel stop or standby point. Location updates are transmitted only when a vehicle starts to move or every $15$ seconds while moving. Speed data is encoded in $5\:\mathrm{mph}$ increments and heading is transmitted in $15$ degree increments. Approximately $95\%$ of all data traffic is received at LAS Headquarters within $1$ second of transmission. Re-transmissions generally account for less than $1\%$ of the total data volume. 

\subsection{Ambulance vs. Civilian Traffic}
Patterns of ambulance movement differ from civilian traffic mainly due to the fact that ambulance crew travelling with blue lights and sirens on are by law exempt from traffic regulations \cite{uk_law_commission_road_1984} that would otherwise impede progress to a patient. For example, ambulances responding to a call on blue lights are permitted to treat red traffic lights as a give way sign, are able to pass on the wrong side of a keep left bollard and can exceed the speed limit. To illustrate this point, we use the Google Maps Distance Matrix API (MDM) \cite{google_google_2017} as a typical example of civilian traffic models in terms of its route selection strategy and estimation expected arrival times. A comparison of AEU trips recorded in the LAS dataset against estimates obtained by MDM in Figure \ref{fig:google_13} suggests a tendency of the latter to overestimate trip duration by a factor of $1.4$. Similar calculations for FRUs (not shown in the figure) suggest a tendency for MDM to overestimate trip duration by a factor of $1.5$. 

\Figure[!t]()[width=0.45\textwidth]{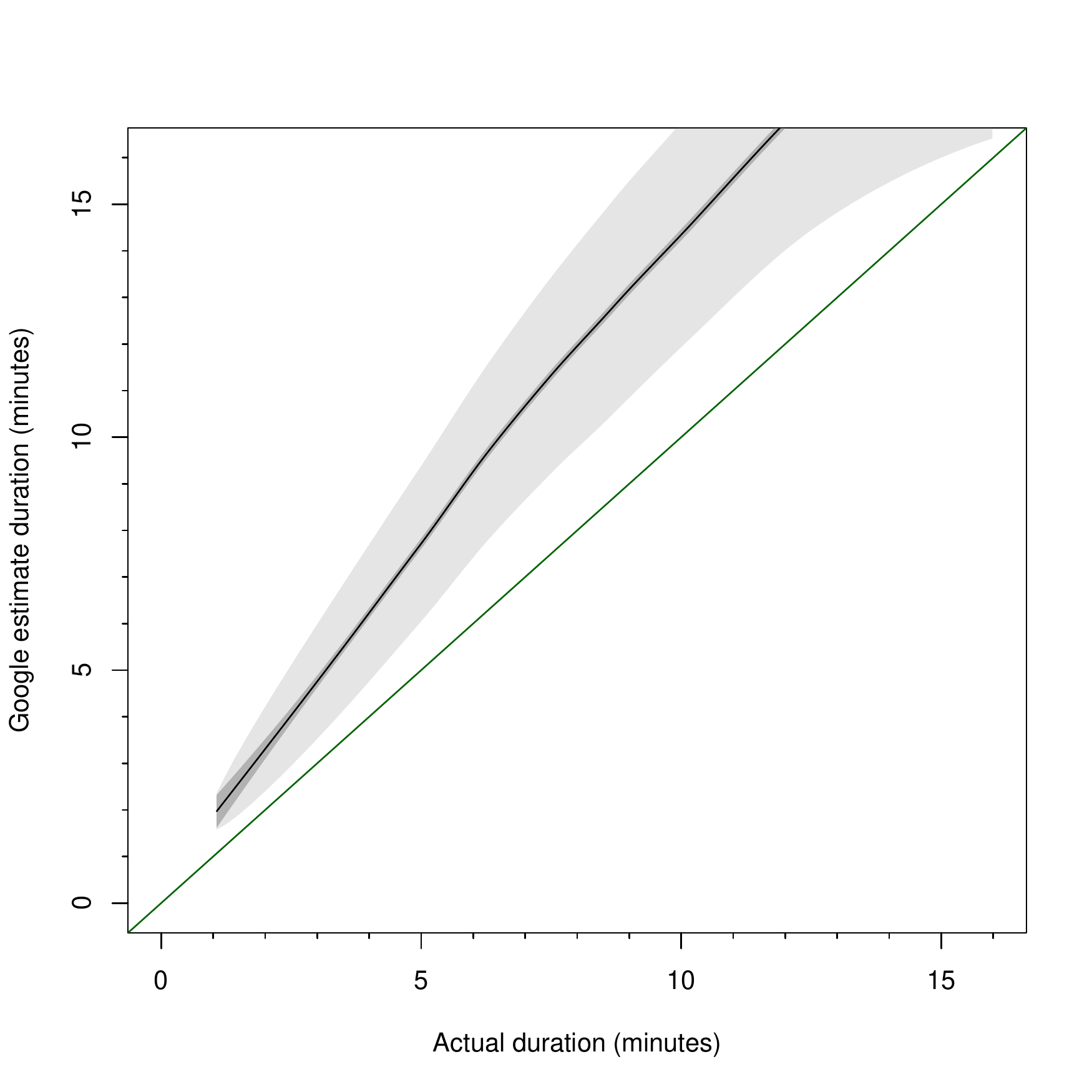}
{Google MDM API trip duration estimates compared against ambulance journey times shown with a $95\%$ probability ribbon. 	\label{fig:google_13}}

Moreover, ambulances travel for a specific purpose that is, in response to emergency medical incidents and the temporal and spatial characteristics of such events also follow particular patterns. Such patterns follow the temporal rhythms of urban life for example, with commuters flowing from the suburbs into the City of London in the morning and returning to their homes in the evening. These diurnal variations in speed of ambulances are observed empirically in our dataset as shown in Figure~\ref{fig:temporalroadspeed}. Our analysis shows that  demand for the service concentrates around several urban nuclei within the metropolitan area of the city reflecting London's well known polycentric urban structure~\cite{10.1371/journal.pone.0015923}, with medical emergencies more likely to occur at specific times and places in London as depicted in aggregate in Figure \ref{fig:incidents_1.png}. 

\Figure[!t](topskip=0pt, botskip=0pt, midskip=0pt)[width=0.35\textwidth]{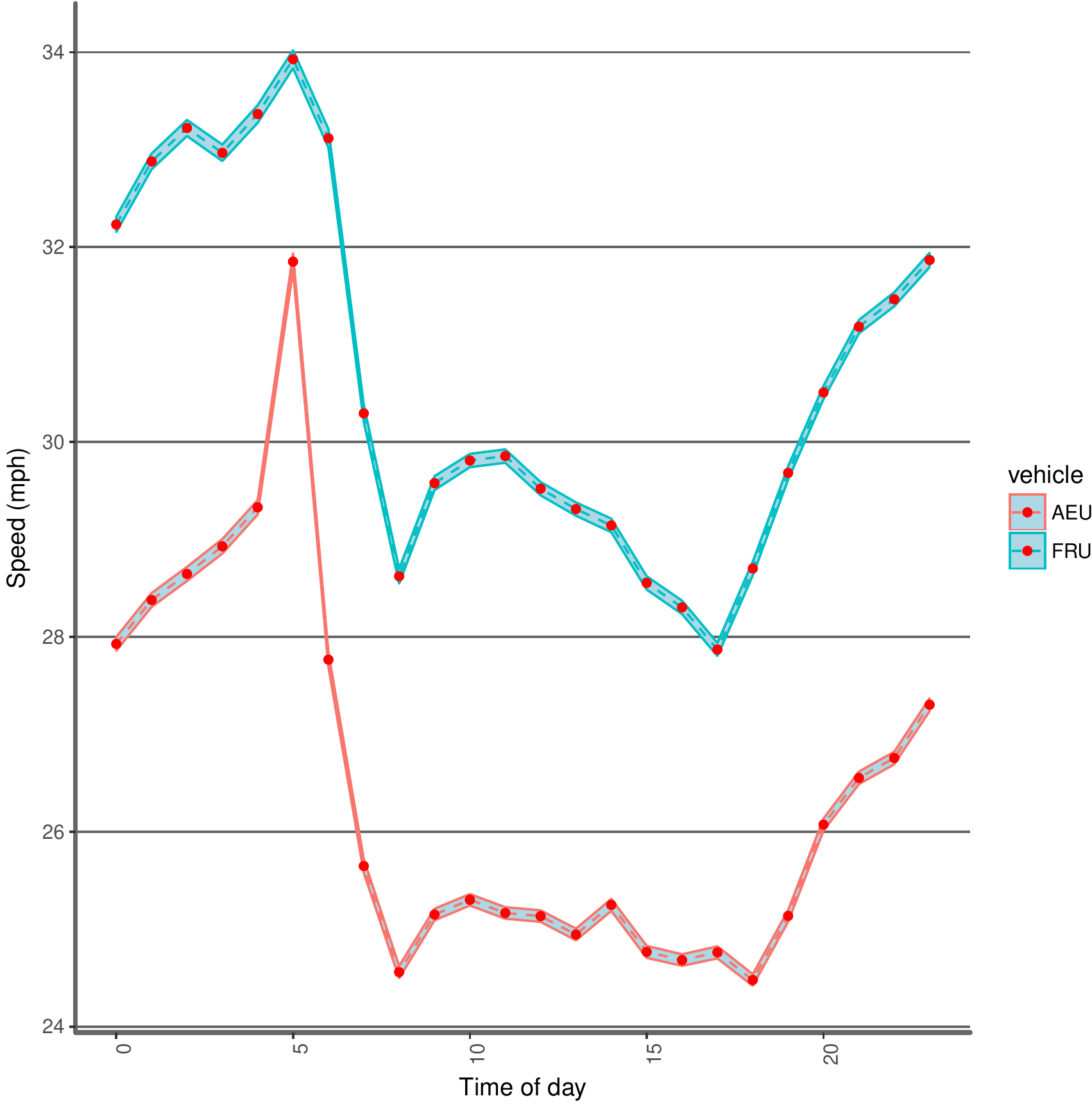}
{Diurnal variation of average road speed for AEU and FRU vehicles correspondingly - Average speeds vary considerably throughout the day, especially in the early morning rush hour. 	\label{fig:temporalroadspeed}}

\Figure[!t]()[width=0.65\textwidth]{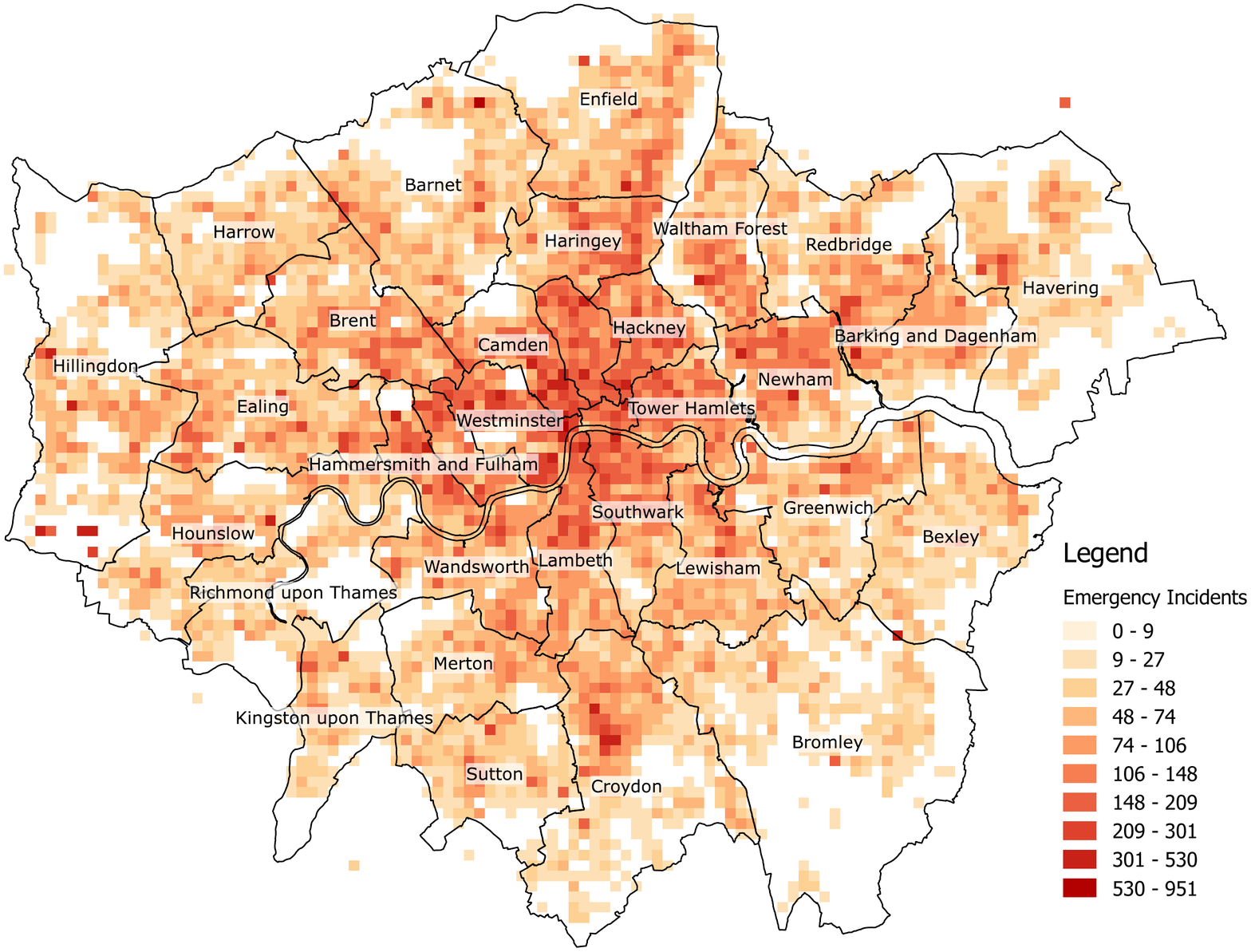}
{Medical emergencies are more likely to occur at specific locations in London reflecting population density 
(showing incident data between September and November 2017). \label{fig:incidents_1.png}}

\subsection{Related Work}

The study of computational techniques for the effective and efficient management of Emergency Medical Service (EMS) resources has a relatively long history \cite{chaiken_methods_1972}. The main focus of research in this domain has been on the development of models for the placement of EMS facilities, such as ambulance stations, and on strategies for resource relocation so that specific performance metrics are maximised \cite{brotcorne_ambulance_2003, daskin_what_2008}. The survey \cite{smith_locational_2009} provides a chronological perspective on the development of research in this area. 

Over the past decade, the wider availability of performance data made available by EMSs worldwide have enabled the exploration of data-driven approaches, which aim to improve response efficiency by adapting to the dynamics of incident generation \cite{Shao, Zhou} and road traffic \cite{Ong2}, leading to a closer focus on improved clinical outcomes rather than resource allocation optimization\cite{mccormack}. Research in this vein has revealed key limitations in historical dispatch strategies, for example the disadvantages of the common practice to send the proximal available ambulance to the incident when choosing the vehicle to dispatch  \cite{Amorim}. Indeed, recent work suggests that EMS systems can benefit from performance improvements achieved through the implementation of sophisticated strategies which take into consideration spatial and temporal factors \cite{Bandara}. 


To account for temporal and spatial variations in resource availability and incidence generation, a common approach is to employ synthetic models for event-driven simulation \cite{anagnostou_distributed_2017, wu_c_using_2009, buuren_simulation_2015,fakhimi_operations_2013}. Nevertheless, when considering the route selection problem specifically, current literature typically adopts a traditional optimisation approach ~\cite{parragh2009ambulance, nordin2012finding, talarico2015ambulance, panahi2009dynamic, nordin2011application} such as linear programming with constraints using tree-based search for shortest path calculations. 

This paper adopts a data-driven methodology for the estimation of ambulance routes and arrival time at the location of an incident as a core ingredient of our approach. The development of such a data-driven methodology has become possible due to the availability of a detailed and comprehensive dataset from LAS recording the true mobility patterns of ambulances at the individual level over a prolonged period of time rather than in the aggregate. To the best of our knowledge this is the first time that a fundamentally data-driven methodology has been applied for the algorithmic inference at high fidelity of the route followed by ambulances responding to emergency incidents with blue lights and sirens on, and the estimation of their arrival time. 

\subsection{Methodological Approach}

Our approach involves estimating the journey time for an emergency ambulatory vehicle while travelling with blue lights and sirens on. This calculation consists of three steps: First, we use historic data to estimate the average speed at which an ambulance responding to an emergency event traverses each road segment in London. Second, from these average speeds we can estimate the journey time for any  route by accumulating the estimated journey time for each road segment in that route. This estimate may include a correction depending on either the number of turns or the length of the route.  Finally, a standard graph theoretic algorithm is used to determine the quickest route between any start and end locations notably from any position of ambulance activation to the location of an incident. 

Central  to  our approach is the computation of the so-called {\it Blue Lights Road Network} (BLRN) for London. It differs from the standard road network used for civilian traffic in that it incorporates the exceptions established by law for emergency vehicles such as, passing on the wrong side of a keep left bollard and use of bus lanes. Moreover, the BLRN incorporates five layers used to associate a road segment with five alternative estimates of average observed speed calculated from the LAS dataset, see next paragraph. The BLRN underpins all our subsequent inference.

The first step, estimating average road segments speeds, is not straightforward. The historic LAS dataset contains GPS locations and vehicle identifiers, this data needs to be matched to the road network in order to infer the route taken by a particular vehicle. In addition, the speeds of emergency vehicles travelling with blue lights and sirens on show hourly and weekly seasonality. Considering the size of the BLRN this implies that our historic data, although substantial, does not provide complete coverage.  Consequently, five different methods for estimating the speed, denoted Metrics I to V, are investigated. These methods are broadly ordered by increasing spatio-temporal granularity, full details are provided in Section \ref{section:roadnetwork}. Metrics I and II are the most straightforward; Metric I assumes that the estimated  average speed is the same over the entire network and Metric II that an already known speed profile exists for each road type.  Metrics III to IV estimate average speeds for individual road types with increasing temporal granularity. Metric V is similar to Metric IV but at higher spatial granularity, considering individual road segments, and includes a more sophisticated method for filling-in missing data.

We show that estimating the entire journey time by simple aggregation will underestimate the result for Metric V, a simple correction is proposed and its effect is described. We also explore the advantages and limitations of different route selection strategies employing  Metrics I to V against both {\it route similarity}, that is the ratio of overlap between the estimated and observed paths; and, {\it arrival time error}, that is the difference in estimated and actual arrival times. Finally, we construct a so-called hybrid model that provides the best match for the routes selected by the current LAS dispatch system and LAS ambulance crews. The hybrid model employs a combination of Metric II with Nelder-Mead optimised road speeds for route selection and Metric V for arrival time estimation.

\section{LAS Dataset}
\label{sec:ambulancemobility}

In this section, we introduce the datasets used in this paper: First, we describe the operational data obtained from LAS with particular emphasis on location tracking. We then proceed with some preliminary observations relating to mobility patterns discovered in the data.

\subsection{ Dataset Features}
The LAS dataset contains incident, activation and tracking information of their entire fleet between March 2014 to December 2016 (cf. Table \ref{tab:recordcounts}). Specifically, the following data type records are included:
\begin{itemize}
	\item \emph{Automatic Vehicle Location (AVLS):} Positioning data obtained from the on-board MDT of an AEU or GPS data from an FRU.
	\item \emph{Incident:} Emergency Event data (one record per incident) including time, geographic coordinates and urgency status.
	\item \emph{Activation:} One activation record per vehicle dispatched incorporating the time and vehicle location at the point of activation and the arrival time at the place of the incident.
\end{itemize}

\begin{table}[!t]
	\centering
	\caption{Key characteristics of the LAS data set. BL indicates that the data relate to travel with blue lights and sirens on. }
	\label{tab:recordcounts}
	\vskip .5cm
	\begin{tabular}{ | l | r | c | c | }
		\hline
		\textbf{Dataset} & \textbf{No. Records} & \textbf{From Date} & \textbf{To Date} \\ \hline
		AVLS & 392,579,544 &	2014-03-01 & 2016-12-31 \\ \hline
		Activations & 6,199,278 &	2011-01-01 & 2016-01-01 \\ \hline
		Incidents &	7,078,201	& 2011-01-01 & 2016-11-30  \\ \hline
		BL AVLS & 71,542,467 &	2014-03-01 & 2016-12-31 \\ \hline
		BL Journeys & 2,311,661 &	2014-03-01 & 2016-12-31 \\ \hline
		BL Incidents & 1,367,649 &	2014-03-01 & 2016-12-31 \\ \hline
	\end{tabular}
	
\end{table}

A regular flow of data was received during the period of data collection with the notable exception of May 2014 when significantly higher volumes were observed. This behaviour was validated against LAS archives which record the weekend of 17-18th May as one of the busiest since records began, with Sunday 18th May being the sixth busiest day in its history. Further, since we are only interested in ambulance mobility when travelling with blue lights and sirens on, AVLS data were filtered for records corresponding only to vehicles en route to a Category A incident, retaining only the relevant data points as summarised in Table \ref{tab:recordcounts}.


\subsection{Location sampling}
The frequency of AVLS reporting was analysed finding that the majority of data arrive at $15$-second intervals as expected. However, a small proportion of records appear to arrive at $0$, $10$ and $20$-second intervals. Further investigation revealed that this was due to erroneous time-stamping by the MDT. Specifically, the MDT is programmed to poll the SNU only once every $15$ seconds so that when it subsequently transmits status updates, for example in response to an action by a crew member, ambulance position is reported using the last cached entry obtained so that stale GPS position information is used.  AVLS records with such erroneous time-stamps were identified and excluded from the dataset.


\subsection{Trace Aggregation}
The first step in the generation of road routes from raw GPS data, is the aggregation of AVLS records into tracks representing individual journeys. This is achieved by grouping AVLS data by call sign, incident identifier, and vehicle type. Note that the call sign is a code used on the radio for the purposes of crew identification. This process generated approximately $2.3$ million distinct journeys attending to approximately $1.3$ million emergency incidents (cf. Table \ref{tab:recordcounts} for details). The higher number of journeys reflects the fact that often multiple vehicles attend a single emergency. 

\subsection{Road network coverage}
For each AVLS data point, we determine the nearest link on the road network using a na\"ive algorithm often referred to as GPS snapping. Snapping each GPS position to its nearest road link provides a simple way to determine the proportion of the road network that is covered by the data set as demonstrated in Figure \ref{fig:uniqueroadlink}. This calculation also enables us to estimate that the entire London road network is fully covered approximately every $2-3$ years suggesting a rough measure of the period required to obtain a full dataset refresh.

\Figure[!t]()[width=0.35\textwidth]{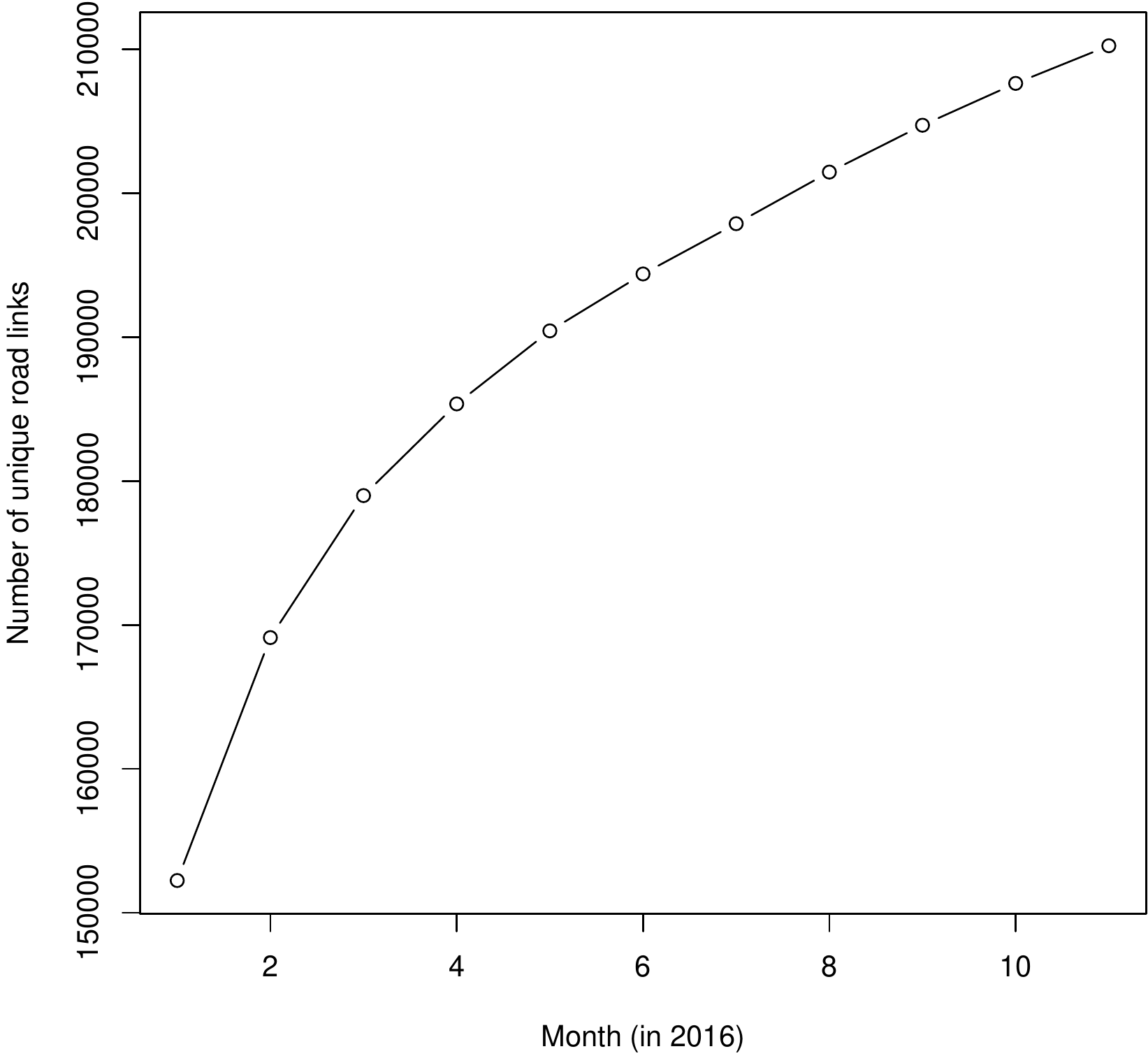}
{Cumulative number of unique road links transversed per month obtained through naive road snapping of AVLS data.\label{fig:uniqueroadlink}}

\subsection{Patterns of ambulance mobility}

Further to the observation that ambulance speed fluctuates with the time of day (cf. Figure \ref{fig:temporalroadspeed}), we note that average speeds range between $24-32\;\mathrm{mph}$ for AEUs and between $28-34\:\mathrm{mph}$ for FRUs. Traffic slows down during the morning (06:00-09:00) and evening rush-hour (16:00-19:00). Day-time speeds improve somewhat from that minimum, with significant further increases during the night. While qualitatively the diurnal pattern is similar for both vehicle types, as expected FRUs are on average quicker by $2-5\;\mathrm{mph}$. 

Considering the spatial distribution of emergency response trips, Figure \ref{fig:spatial-7_.pdf} depicts their density at $3\:\mathrm{km}\times 3\:\mathrm{km}$ cell resolution. As expected, Figure \ref{fig:spatial-7_.pdf} confirms significant differences between centre and suburbs. It also reveals considerable localised differences in the outer regions in particular with several hot-spots of higher average incident density identified roughly corresponding to suburban loci.

\Figure[!t]()[width=0.45\textwidth]{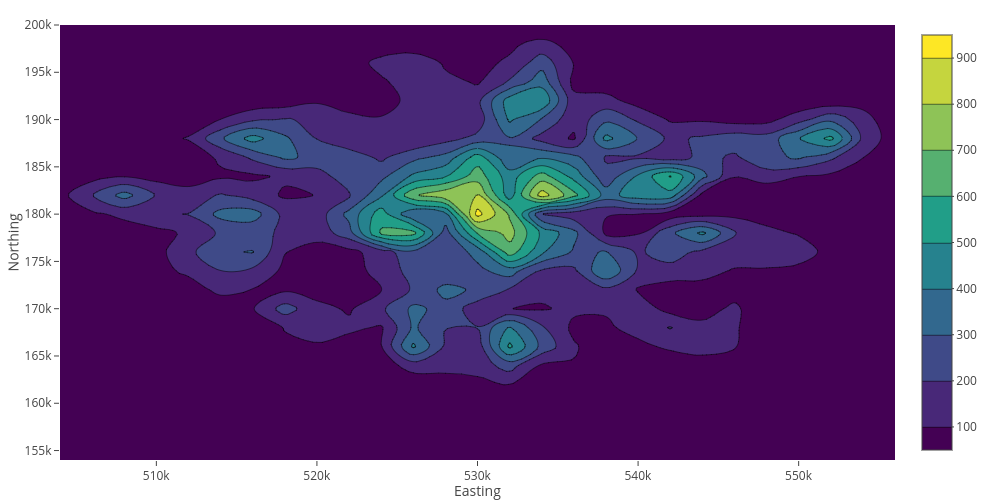}
{Number of journeys in each cell across London where each cell is 3km x 3km. The outer regions of the city have far fewer journeys than the inner core. \label{fig:spatial-7_.pdf}}


Overall, these observations suggest that an effective model for ambulance route selection and arrival time prediction should be adaptive to the spatial and temporal context and take into account the specific vehicle type involved. Indeed, the approach adopted in this study and detailed next is motivated by this observation. 

\section{Blue Lights Road Network}\label{section:roadnetwork}
A core data structure employed in the present work is the multi-layer directed graph $G$ representing London's road network as viewed by emergency vehicles while travelling with blue lights and sirens on. $G$ forms the foundation for modelling ambulance vehicle mobility and  the evaluation of alternative route selection methods. Specifically, the Blue Lights Road Network (BLRN) $G$ is the graph: 
\begin{equation*}
G^{(j)}=\{N, E, W^{(j)}\}, \quad j=\RN{1}, \RN{2}, \ldots, \RN{5},
\end{equation*}

such that:

\begin{itemize}
\item  the nodes $N$ correspond to road intersections,
\item the edges $E$ are road segments connecting intersections (we refer to edges also as road links), 
\item each road link $e_i\in E$ is associated with weights $W^{(j)}_i$ for $j=\RN{1}, \RN{2}, \ldots, \RN{5}$  corresponding to features related to the estimated speed of an ambulance travelling across $e_i$ under Blue Lights using five alternative speed metrics. 
\end{itemize}
The calculation of $W^{(j)}$ for each road link is a key element of the present work. Specifically, we consider five alternative road link cost-estimation metrics of increasing sophistication:
\begin{itemize}
\item Metric I uses the fixed speed of $22.8\;\mathrm{mph}$ for all road links. This metric corresponds to the na\"ive approach assuming there are no differences in speed ratios and traffic conditions across the city. It is only used in this paper as a baseline for comparison. 

\item Metric II uses a standard speed profile for each road type (cf. Table \ref{tab:road_speed_las_only}) also adding a delay of $2.5$ seconds each time a junction is crossed. Metric II corresponds to the method employed by the LAS routing engine at the time when the data set was recorded. 

\item Metric III uses road speed estimates adapted to ambulance position, hour-of-day and ambulance type. Specifically, the road link $e_i$ is associated with the $2\times 24$ weight matrix $W^{(\RN{2})}_i$, with rows corresponding to either AEU or FRU and columns to hour-of-day. The weight $W^{(\RN{2})}_i(a,h)$ for a specific ambulance type $a$ during hour-of-day $h$ are calculated as the harmonic mean of all AVLS records in the data set located within a $500\times 500\mathrm{m}$ box centred at the midpoint of $e_i$ recorded on roads of the same type during hour-of-day $h$ (i.e. irrespective of the day of the week).

\item Metric IV uses road speed estimates characteristic of ambulance position and ambulance type similar to Metric III, but adapting to the hour-of-week. Hence, in this case the weight matrix $W^{(\RN{4})}_i$ has size $2\times 168$ and calculated in a manner similar to Metric III incorporating variations in weekly traffic conditions. Metric IV improves on Metric III in terms of time resolution and the ability to capture weekly patterns, but its accurate calculation requires a significantly higher data set size to provide an adequate number of samples for each $e_i$ in the road network.

\item Finally, Metric V estimates the weight matrix $W^{(\RN{5})}_i$  for road link $e_i$ as the harmonic  mean of all speed estimates recorded on $e_i$ per  hour-of-week and vehicle type. If no records have been recorded on the specific road link $e_i$ then Metric IV is employed instead.  
\end{itemize}

\begin{table}[!ht]
	\centering
	\caption{Speed by road type used by the LAS routing engine (in $\mathrm{mph}$).}
	\label{tab:road_speed_las_only}
	\vskip .5cm
	\begin{tabular}{|l|r|}
		\hline 
		\textbf{Road type} & \textbf{Speed} \\ 
		\hline 
		Motorway					& 35 \\	\hline 
		A Road					& 29 \\ \hline 
		B Road					& 24 \\	\hline 
		Minor Road				& 19 \\	\hline 
		Local Street				& 14 \\	\hline 
		Private Road- Publicly 		& 5 \\	\hline 
		Private Road- Restricted		& 5 \\	\hline 
		Alley						& 3  \\	\hline 
		Pedestrianised Street		& 2 \\	\hline 
	\end{tabular} 
\end{table}


To construct the BLRN we employed the Integrated Transport Network (ITN) dataset supplied by Ordnance Survey. The ITN was modified to allow routing according to the rules pertaining to ambulances and other emergency vehicles travelling with blue lights and sirens on. Specifically, the BLRN permits:

\begin{itemize}
\item passing on the incorrect side of \textit{Keep Left/Right Signs} including passing on the wrong side of a keep left bollard,
\item right turns where \textit{No Right Turn} restrictions apply,
\item use of bus lanes during operating hours,
\item driving into a pedestrian precinct,
\item treating red traffic lights and zebra crossings as a give way sign, and
\item  exceeding the speed limit.
\end{itemize}
Moreover, the BLRN replaces single edges for bi-directional roads with two edges, one for each direction. This modification is required for the effective implementation of the map-matching methods introduced in Section~\ref{section:map-matching}, which represent a key ingredient of our data-driven methodology. Finally, the estimation of the weights $W^{(i)}$ requires the calculation of ambulance speed for each road segment in every journey reconstructed from the AVLS records. We elaborate further on this point in Section~\ref{section:map-matching}. 

\section{ Ambulance Route Reconstruction}
\label{section:map-matching}
As already noted, the accurate reconstruction of observed ambulance routes provides the foundation for the development of alternative road speed models for ambulances travelling under blue lights and sirens on. Because LAS ambulances record their position relatively infrequently while operating in a high-density urban setting, the GPS tracks obtained from AVLS records provide only a coarse-grain record of their movement. In particular, such tracks do not identify all the road links of the route followed even after snapping to the road network, which in turn severely limits our ability to produce accurate speed estimates. To address this problem, we adopt a map-matching approach \cite{hashemi_critical_2014} that leads us to achieve the full reconstruction of ambulance routes from relatively low-frequency GPS traces. Overall, this approach enables significant improvements to road speed models using Metrics III, IV and V.

\subsection{Map-matching AVLS tracks}
The map-matching process transforms a time-ordered sequence of GPS locations into the most likely path followed by the ambulance in the BLRN. We investigated the performance of several map-matching techniques based on particle filters and the Hidden Markov Model/Viterbi Paths (HMM/V) method on the LAS dataset specifically. This was deemed necessary due to the fact that each technique provides different trade-offs including sensitivity to the sampling rate, computational complexity, real-time performance and accuracy. 

Two popular map-matching methods were considered in detail, one employing particle filters \cite{p._davidson_application_2011, par_parallelization_2011, mokhtari_map_2014} and the second the HMM/V variant proposed by Newson and Krumm \cite{newson_hidden_2009}. Our experiments suggest that AVLS data are best suited to the HMM/V approach which we found to produce consistently higher fidelity routes (the details of these experiments are not included in this paper due to lack of space). Nevertheless we note that this result is consistent with the literature \cite{wei_fast_2012, lou_map-matching_2009-1, jagadeesh_online_2017, luo_enhanced_2017} in that it confirms that HMM/V produces accurate results for the relatively low sampling rate such as one AVLS record every $15$ seconds typical in the LAS dataset; it does not require setting tight speed limits, which fits well with the fact that ambulances are exempt from such limits when travelling with blue lights and sirens on; and, adapts well to low accuracy GPS fixes which can occur especially in urban settings. 


Using HMM/V map-matching, we obtain the complete road segment-by-segment route followed by an ambulance and the time of entry to and exit from each road segment. Assuming constant travelling speed across individual road segments, vehicle speed can be estimated directly as the ratio of road segment length over traversal time. 

\subsection{BLRN Coverage}
A total of $1,910,941$ journeys were processed by the HMM/V algorithm involving $177,975,172$ road link records and associated speeds. 
Using the fully reconstructed routes, we carry out an analysis of coverage by road type: Table \ref{tab:snaptablehmm} suggests that coverage for road types more likely to be used by an ambulance is considerably improved against GPS snapping. For example, for Type A roads (major roads providing large-scale transport links) coverage increases to $94\%$. 

Table \ref{tab:road_speed_las} compares road speed estimates obtained from GPS against those calculated after map-matching per road type and the fixed road speeds used by the routing engine implemented by LAS as the baseline. Overall, these results highlight the significance of map-matching, which yields considerable improvements. 

\begin{table}[!ht]
	\centering
	\caption{BLRN coverage by road type by HMM/V and GPS snapping. The table shows: a) the total number of road links in BLRN, b) the number of road links visited, c) the proportion of BLRN covered by HMM/V,  and d) the proportion of  BLRN covered by naive snapping. Note that GPS is snapped to a bi-directional road link, whereas HMM/V also identifies the direction of travel.}
	\label{tab:snaptablehmm}
	\vskip .5cm
	\begin{tabular}{|l|r|r|c|c|}
		\hline 
		\textbf{Road type} & \textbf{ Links} & \textbf{ Used} & \textbf{HMM/V } & \textbf{GPS} \\ 
		\hline 
		A Road					& 65,857	& 62,124	& 94\%	& 78\% \\ 	\hline 
		Alley						& 46,473	& 15,579	& 32\%	& 34\% \\	\hline 
		B Road					& 21,747	& 18,257	& 84\%	& 64\% \\	\hline 
		Local Street				& 327,827	& 234,624	& 69\%	& 50\% \\	\hline 
		Minor Road				& 62,052	& 48,853	& 78\%	& 59\% \\	\hline 
		Motorway					& 1,131		& 689		& 61\%	& 90\% \\	\hline 
		Pedestrianised Street		& 363		& 317		& 87\%	& 70\% \\	\hline 
		Private Road- Publicly 		& 6,735		& 3,945		& 56\%	& 55\% \\	\hline 
		Private Road- Restricted		& 75,360	& 34,151	& 7\%	& 45\% \\	\hline 
		\textbf{Total} & \textbf{607,545}	& \textbf{418,539}		& \textbf{69\%}	& \textbf{65\%} \\	\hline 
	\end{tabular} 
\end{table}

\begin{table}[!ht]
	\centering
	\caption{Road speeds by road type a) estimated from map-matching b) recorded from GPS, and c) from LAS routing engine. All speeds are in $\mathrm{mph}$.}
	\label{tab:road_speed_las}
	\vskip .5cm
	\begin{tabular}{|l|r|r|r|}
		\hline 
		\textbf{Road type} & \textbf{HMM/V} & \textbf{GPS} & \textbf{LAS} \\ 
		\hline 
		A Road					& 31.21 & 20.74 & 29 \\ \hline 
		Alley						& 22.51 & 14.61 & 3  \\	\hline 
		B Road					& 28.01 & 18.01 & 24 \\	\hline 
		Local Street				& 18.79 & 10.89 & 14 \\	\hline 
		Minor Road				& 26.80 & 16.69 & 19 \\	\hline 
		Motorway					& 42.65 & 38.79 & 35 \\	\hline 
		Pedestrianised Street		& 16.29 & 8.12  & 2 \\	\hline 
		Private Road- Publicly 		& 16.58 & 8.14  & 5 \\	\hline 
		Private Road- Restricted		& 18.50 & 10.55 & 5 \\	\hline 
	\end{tabular} 
\end{table}

\section{Modelling Ambulance Movement}
\label{section:modelling}

Using the speed estimates calculated in Section \ref{section:map-matching}, we can compute the weights $W^{(j)}$ defined in Section~\ref{section:roadnetwork} for each road segment in the BLRN. Then, for a particular incident and initial ambulance location, we can apply a shortest path algorithm on $G$ using each of the five weight layers  $W^{(j)}$ for $\;j=\RN{1},\RN{2},\ldots,\RN{5}$ in turn to select a route. In this section, we explore the advantages and limitations of these alternatives by comparing them against a testing set selected from the LAS dataset along two criteria: (a) travel time error, and (b) route similarity. Without loss of generality, we use the classic Dijkstra's algorithm for shortest path calculations throughout the remainder of this study.

\subsection{Predicting Arrival Time}\label{Route Selection} 
LAS data up to October 2016 are employed to calculate weights the $W^{(j)}$  and journeys en route to Category A emergencies recorded during November 2016 for testing (approximately $66,000$ routes). First, we consider the precision of estimated arrival times: For each route in the testing set we calculate the difference between the actual journey duration as recorded in the LAS dataset and the estimated arrival time using each of the alternative metrics to select a route. The distribution of the error is depicted at the top of Figure  \ref{ra__ap_all_U.pdf}.


\Figure[!t]()[width=0.35\textwidth]{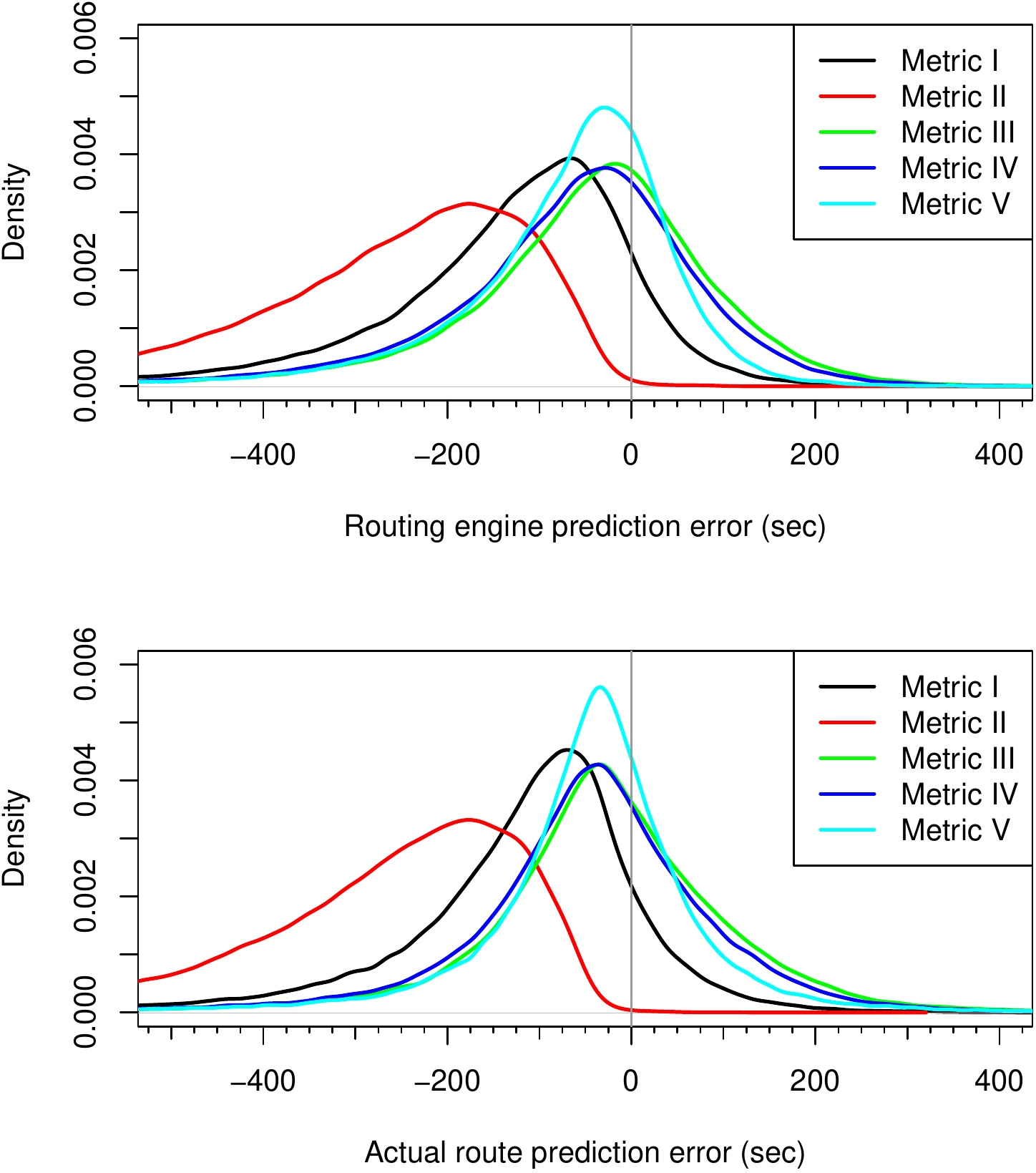}
{Distribution of arrival time error. Top: Weight layers used to calculate both the route and arrival time. Bottom: Weight layers used to calculate only arrival time using the actual route taken by the vehicle as recorded in the LAS dataset. \label{ra__ap_all_U.pdf}}

Figure  \ref{ra__ap_all_U.pdf} suggests that on average all methods tend to underestimate arrival time. For example, using Metric II the error has a mode of approximately $-130$ seconds with $95\%$ of all routes estimated at $-45$ seconds or less than the actual time duration. Surprisingly, Metric I outperforms Metric II even though the latter is used operationally by LAS. The more sophisticated Metric V provides the best performance with mode of approximately $-20$ seconds. 

To better understand the tendency of the models to underestimate arrival times, specifically whether this is due to erroneous route selection or inaccurate speed estimates, we also calculate travel times using each model but using the actual route recorded in the LAS dataset rather than the route selected by the corresponding metric. The outcomes of this comparison are presented at the bottom of Figure  \ref{ra__ap_all_U.pdf} suggesting an improvement and with Metric V still outperforming all others.

\Figure[!t]()[width=0.40\textwidth]{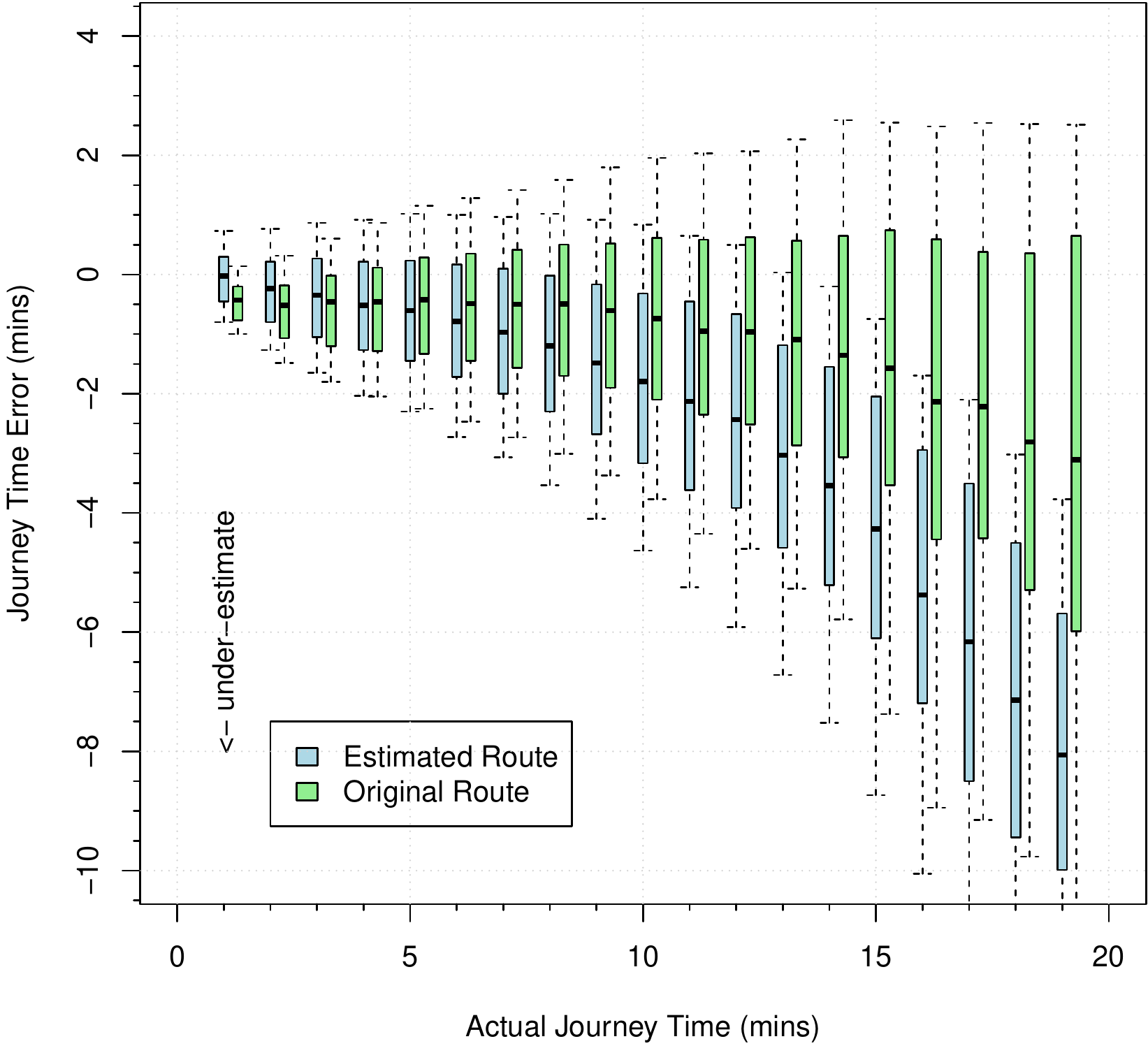}
{Arrival time prediction error for Metric V using actual and predicted routes. Boxes represent 1st and 3rd quartile. Whiskers 0.1 and 0.9 quantiles. 	\label{fig:ra__e_rl_dva2_U.pdf}}


Figure \ref{fig:ra__e_rl_dva2_U.pdf} provides a closer examination of the difference in journey time estimation for the actual (green) and estimated (blue) route taken using only the best performing Metric V. Note that for journeys of up to $8$ minutes there is only a relatively small difference in accuracy. Recall that due to the regulatory framework under which LAS operates, response time of $8$ minutes is required for $75\%$ of Category A incidents resulting in the largest proportion of our dataset to record sub-$8$-minute journeys. For journeys longer than $8$ minutes, arrival time estimates diverge with the mean estimated journey time using the actual route providing good accuracy for up to $20$ minutes. 

\subsection{Route Similarity}
To further investigate the different path choices between predicted and actual routes, we compare routes by segment as shown in Figure \ref{fig:ra__simi-mean_U}. First, each route is split in four equidistant segments $q_1, q_2, q_3$ and $q_4$. Then for each segment we calculate \textit{path coincidence} as the percentage of road links in the actual route that are correctly identified. Path coincidence ranges between 70\% and  80\%. As expected, Metric II used by LAS for route planning, produces the most similar routes as expected (however, recall that this does not result into the most accurate arrival time estimation cf. Figure \ref{ra__ap_all_U.pdf}). Moreover, the more sophisticated Metric V does not perform well in terms of route similarity.

\Figure[!t]()[width=0.35\textwidth]{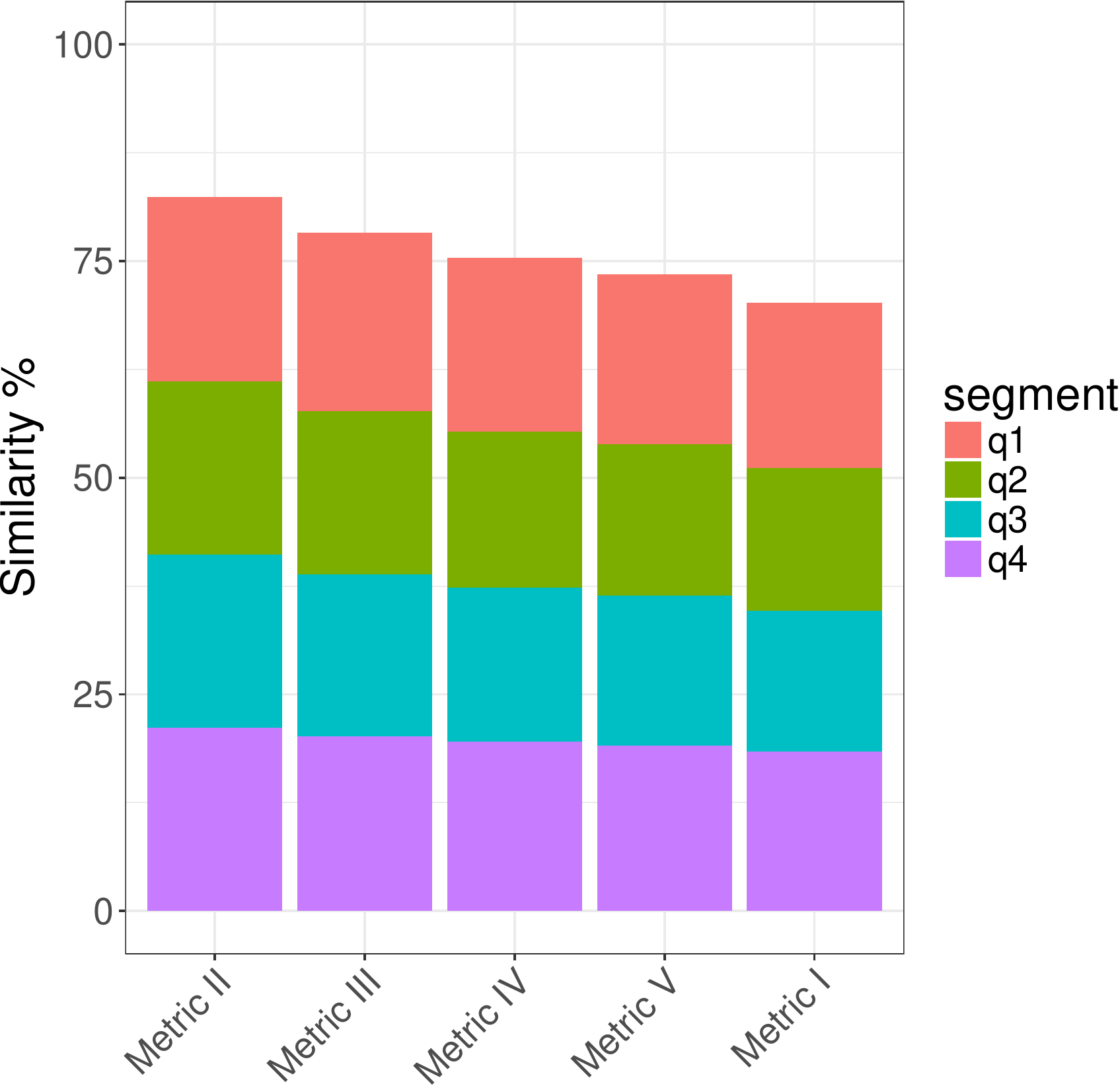}
{Similarity of estimated vs. actual route per route segment. \label{fig:ra__simi-mean_U}}

Figure \ref{fig:ra__simi-rl_U.pdf} shows the distribution of path coincidence for Metric V per route segment: For all segments, scores between $5\%$ and $95\%$ appear fairly uniformly distributed. Further, in segments $q_1$ and $q_4$, representing the beginning and the end of the route, a peak is observed at $100\%$ path coincidence indicating that a large proportion of routes have been matched perfectly. In segments $q_2$ and $q_3$ in addition to similar peaks at $100\%$ path coincidence, a secondary peak is observed at $0\%$ suggesting that an entirely different route was selected. Overall, these observations suggest that in a significant proportion of cases, Metric V picks substantially different routes towards the location of an incident. Moreover, these routes lead to considerably shorter arrival times than those suggested by Metric II.

\Figure[!t]()[width=0.35\textwidth]{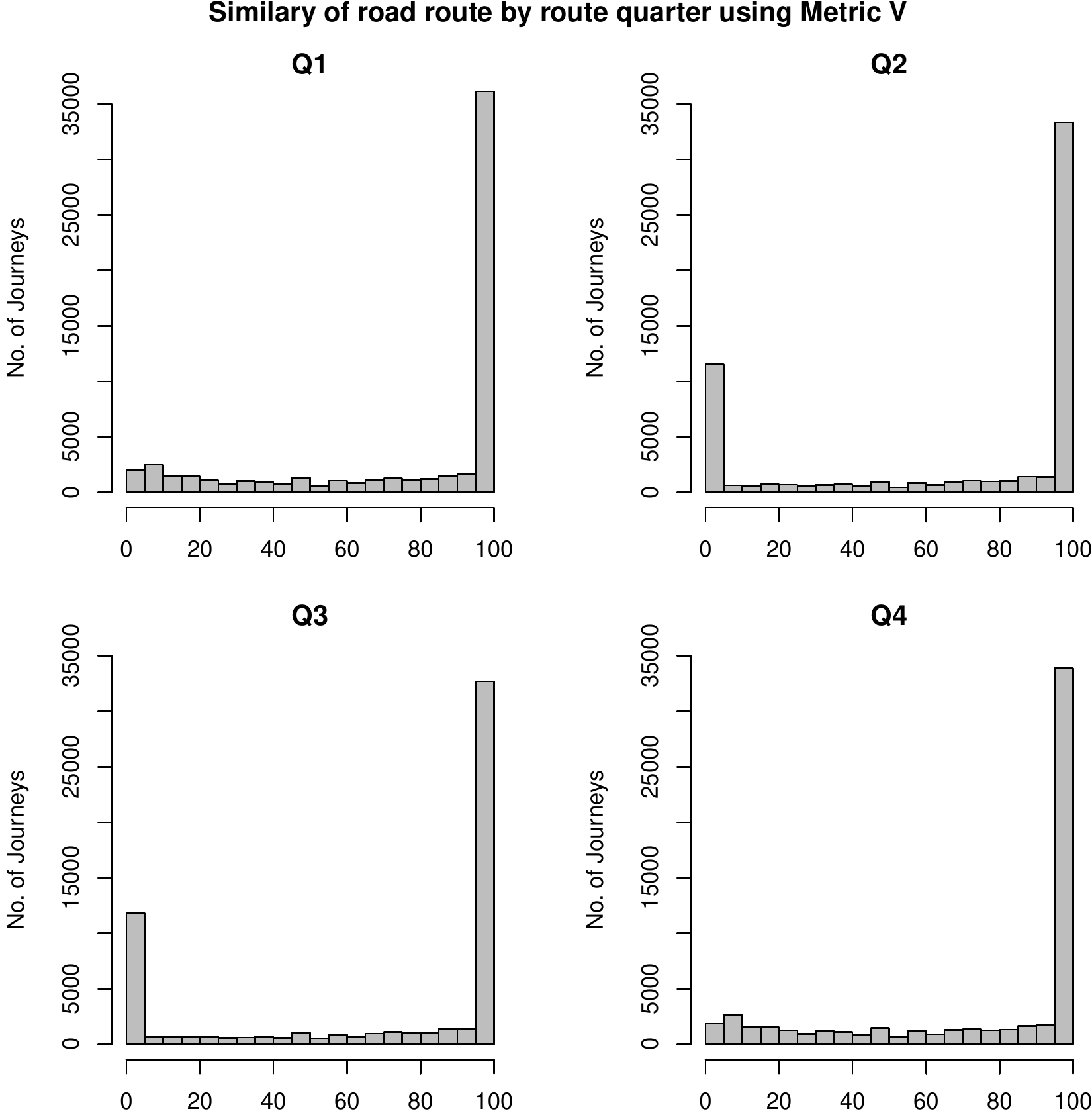}
{Similarity expresses as path coincidence of actual and predicted routes using Metric V per route segment. \label{fig:ra__simi-rl_U.pdf}}

The dataset employed in the present study does not permit further investigation of the causes of the route selection divergence, for example, it does not record how often the crew overrides the route selection made by the MDT rather than following the recommended route. Localised traffic load and related driving conditions are also not recorded, nor are they available via third party datasets to the required resolution. As such, without further experimentation under true emergency response circumstances it is not possible to assess whether the routes selected by the MDT routing engine are optimal specifically for ambulances travelling under blue lights and sirens on. Yet, this observation suggests the possibility that the results presented here offer the opportunity for reduced ambulance arrival times if Metric V were used for route selection. Figure \ref{fig:ra__e_rl_dva2_U.pdf} suggests that the potential improvement is approximately one minute for a ten minute journey, performance which would represent a considerable improvement in the ability of LAS to meet mandated operational targets.

\subsection{Spatio-temporal Variation} 
London is a polycentric, densely populated area with a complex, roughly circular, road network~\cite{10.1371/journal.pone.0015923}. AVLS data confirm that traffic density is higher, and road journeys for comparable distances are thus longer, in the centre of London and certain times of day (cf. Figure \ref{fig:incidents_1.png}). In this section, we consider how our predictions of preferred ambulance routes vary along spatio-temporal dimensions. Specifically, we investigate whether there is a relationship between journey time estimation and distance from the centre of London. To this end, we follow the commonly adopted convention to consider Charing Cross as the centre point of London. We calculate distance from the centre as the length in kilometres of the straight-line segment connecting Charing Cross and the geographic midpoint between start and end locations of the journey.

\Figure[!t]()[width=0.40\textwidth]{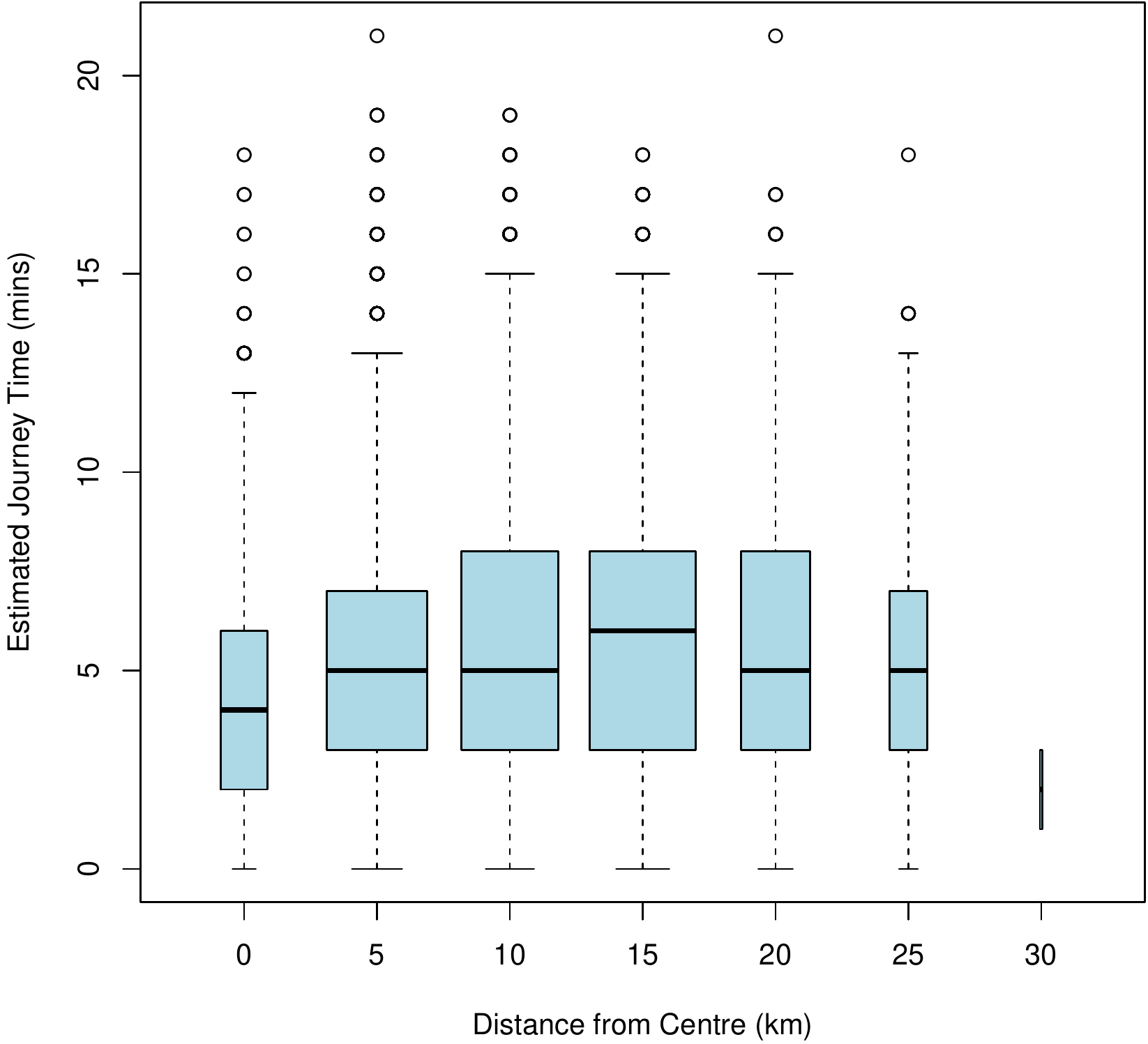}
{Variation of estimated arrival times by distance from the centre of London. \label{fig:ra__dist_acc_U.pdf}}

Figure \ref{fig:ra__dist_acc_U.pdf} shows estimated journey time against distance from Charing Cross. It reveals that journey times are somewhat shorter at the centre but overall roughly consistent at $4-6$ minutes across London. Variation is least at the centre and decreases somewhat at the outer suburbs, but note that the volume of incidents in the furthest areas is low. Because arrival times also depend on the number of resources available, this is also indicative of how close a resource is located to the incident. Finally, Figure \ref{fig:ra__dist_err_U.pdf} considers timing error and path coincidence: In both cases, mean prediction error and variance are higher at the centre, gradually improving further away.  

\Figure[!t]()[width=0.40\textwidth]{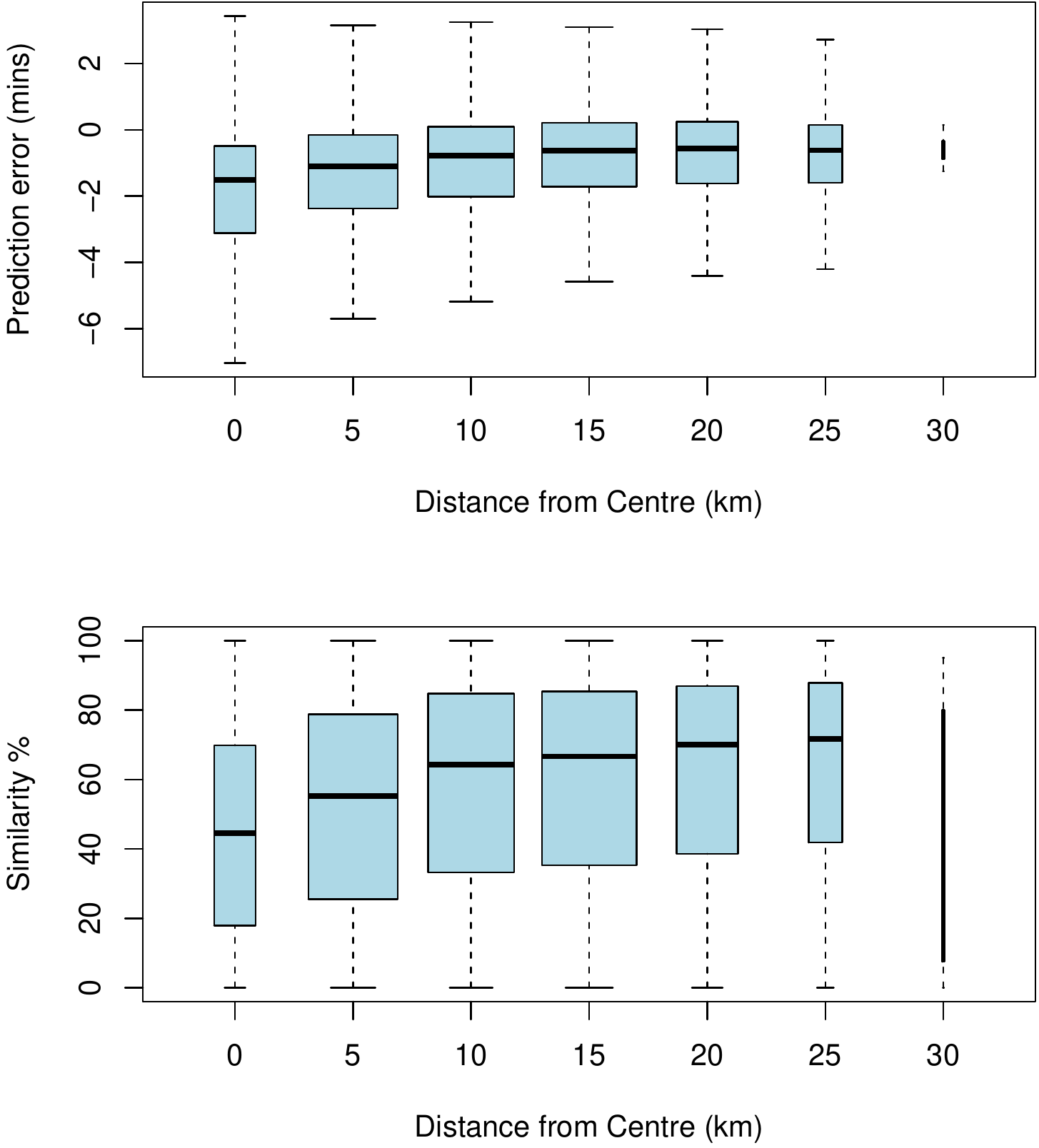}
{Top:  Arrival time error by distance from the centre of London. Bottom: Path coincidence by distance from the centre of London. \label{fig:ra__dist_err_U.pdf}}

Considering temporal variations, Figure \ref{fig:ra__e_rl_hod_U.pdf} displays hour-by-hour error during the day. Performance is relatively uniform irrespective of the time of day with only two noticeable drops, coinciding with the change of crew shifts. 

\Figure[!t]()[width=0.40\textwidth]{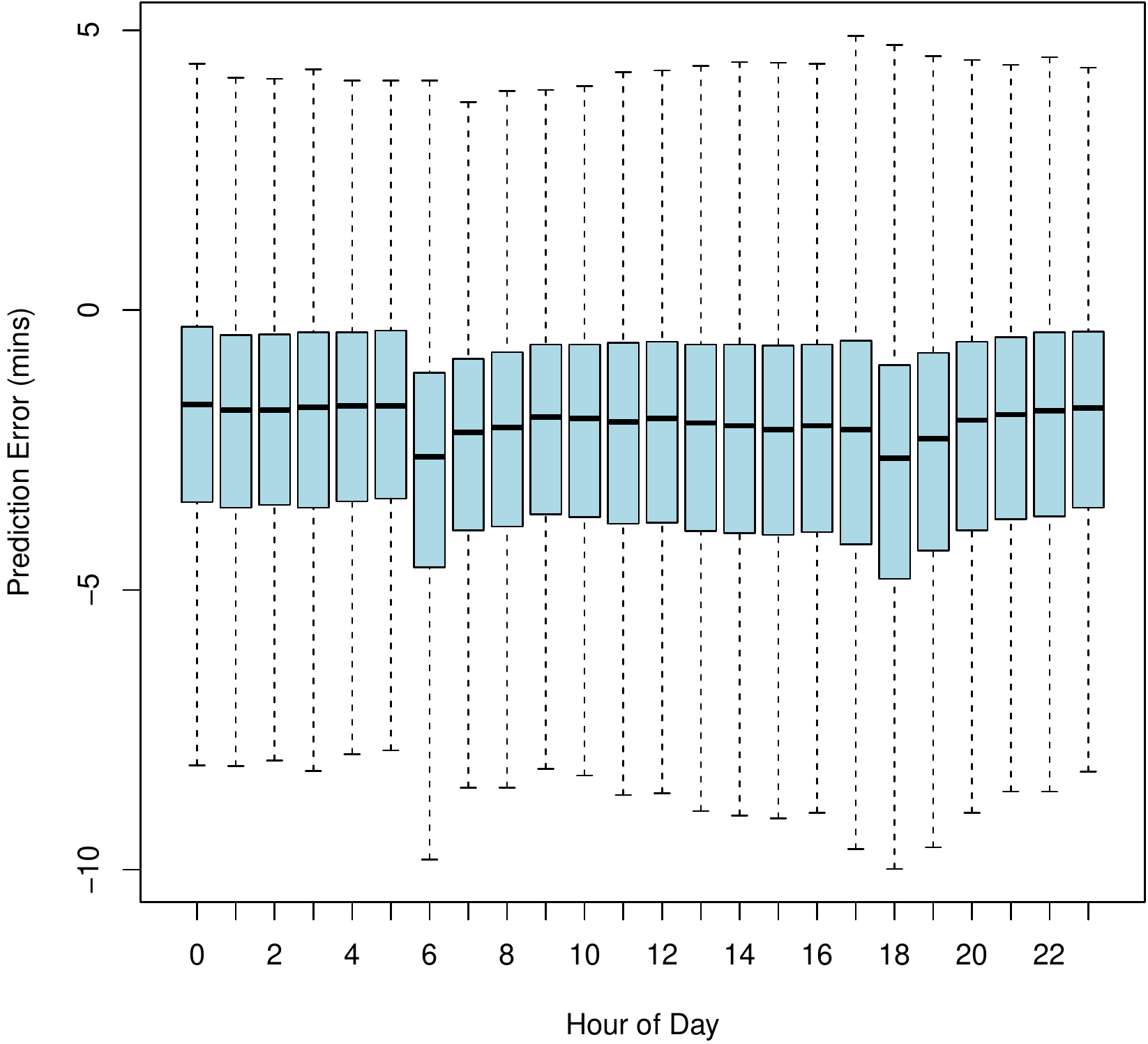}
{Prediction accuracy changes over time using Metric V. - Only small variations occur during the day with dips around shift change. \label{fig:ra__e_rl_hod_U.pdf}}

\section{Hybrid Predictive Model} 
\label{section:hybrid}

In Section \ref{Route Selection} we observed a general tendency to underestimate arrival times in general and when using Metric V in particular. To address this limitation, a na\"ive approach would be to simply adjust arrival times by applying the bias function suggested by Figures \ref{fig:ra__e_rl_dva1_U.pdf} and \ref{fig:ra__e_rl_dva1_CI.pdf}, namely:
\begin{equation*}
t_\chi= \frac{ t_\beta }{0.8029} - 23.3843,
\end{equation*}
where $t_\beta$ is the original estimated journey time and $t_\chi$ is the corrected time in seconds. The result of this adjustment is shown in Figure \ref{fig:ra__e_rl_C.pdf} indicating a significant improvement in accuracy with a mean error of less than $60$ seconds for journeys lasting up to $14$ minutes which account for $90\% $ of all journeys completed by LAS in response to a Category A incident. 

\Figure[!t]()[width=0.45\textwidth]{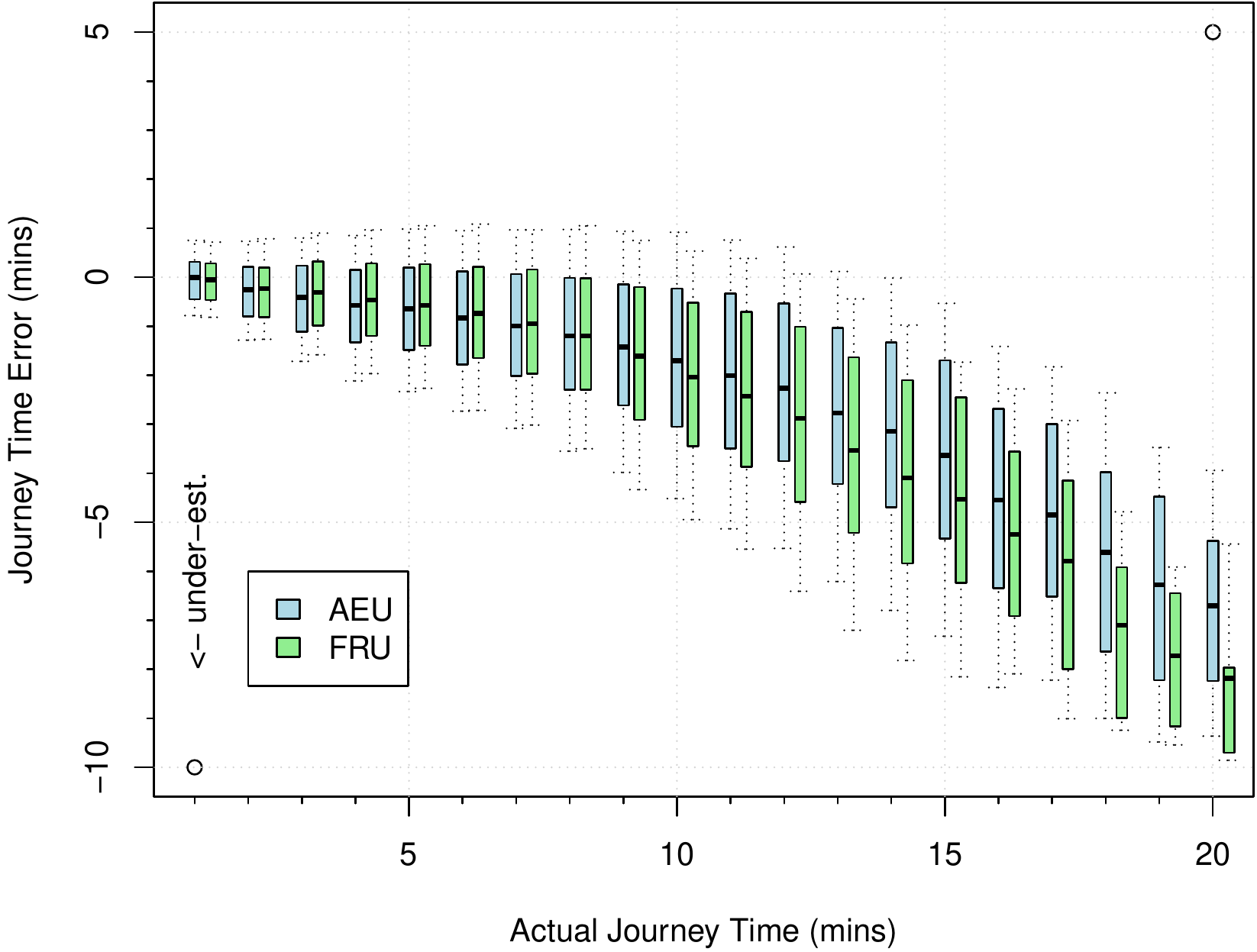}
{Arrival time prediction error using Metric V for AEUs and FRUs. Boxes indicate the 25th to 75th percentiles. Whiskers the 10th and 90th percentiles. \label{fig:ra__e_rl_dva1_U.pdf}}

\Figure[!t]()[width=0.4\textwidth]{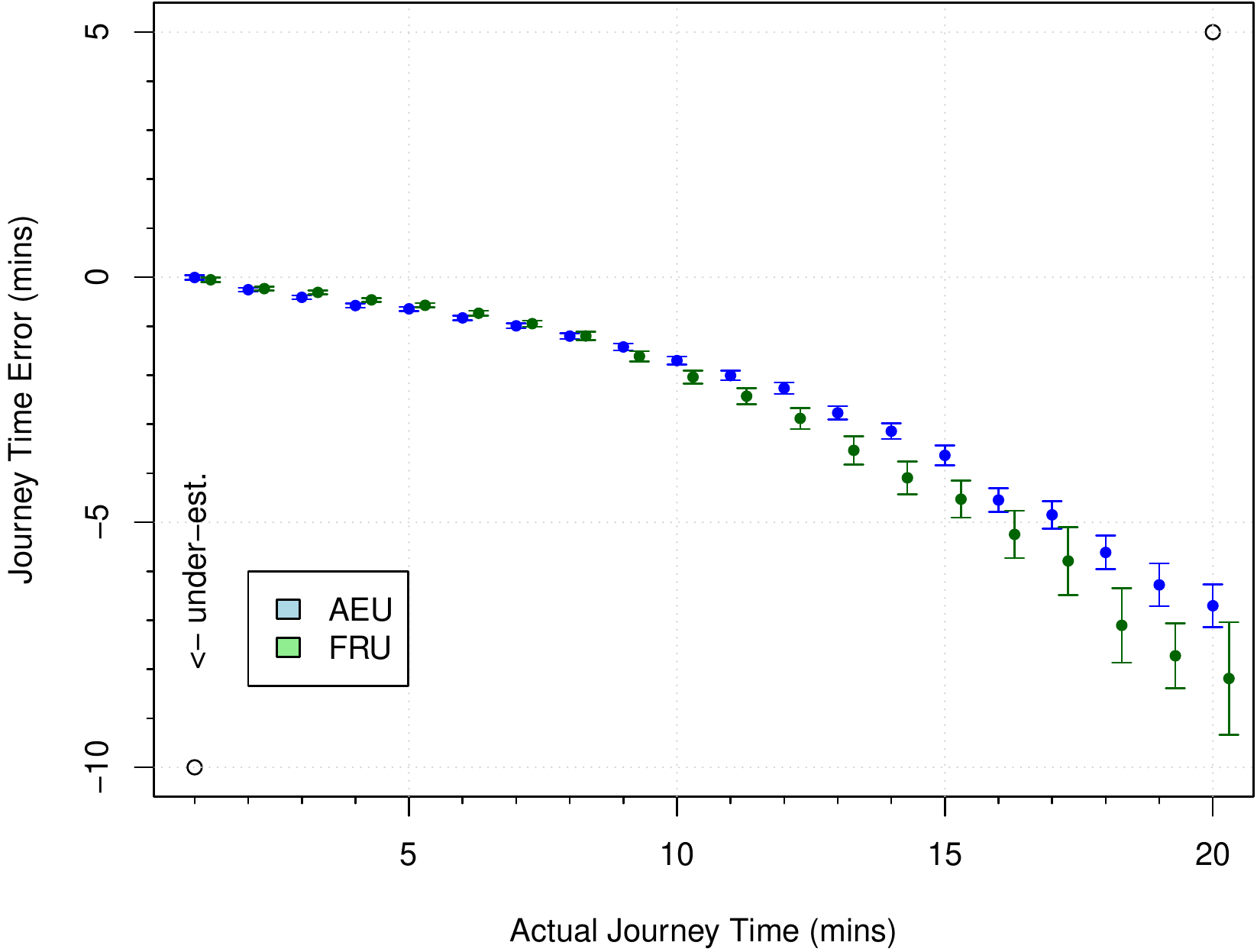}
{Arrival time error prediction using metric V for AEUs and FRUs, showing $95\%$ confidence intervals for the mean. 	\label{fig:ra__e_rl_dva1_CI.pdf}}

\Figure[!t]()[width=0.4\textwidth]{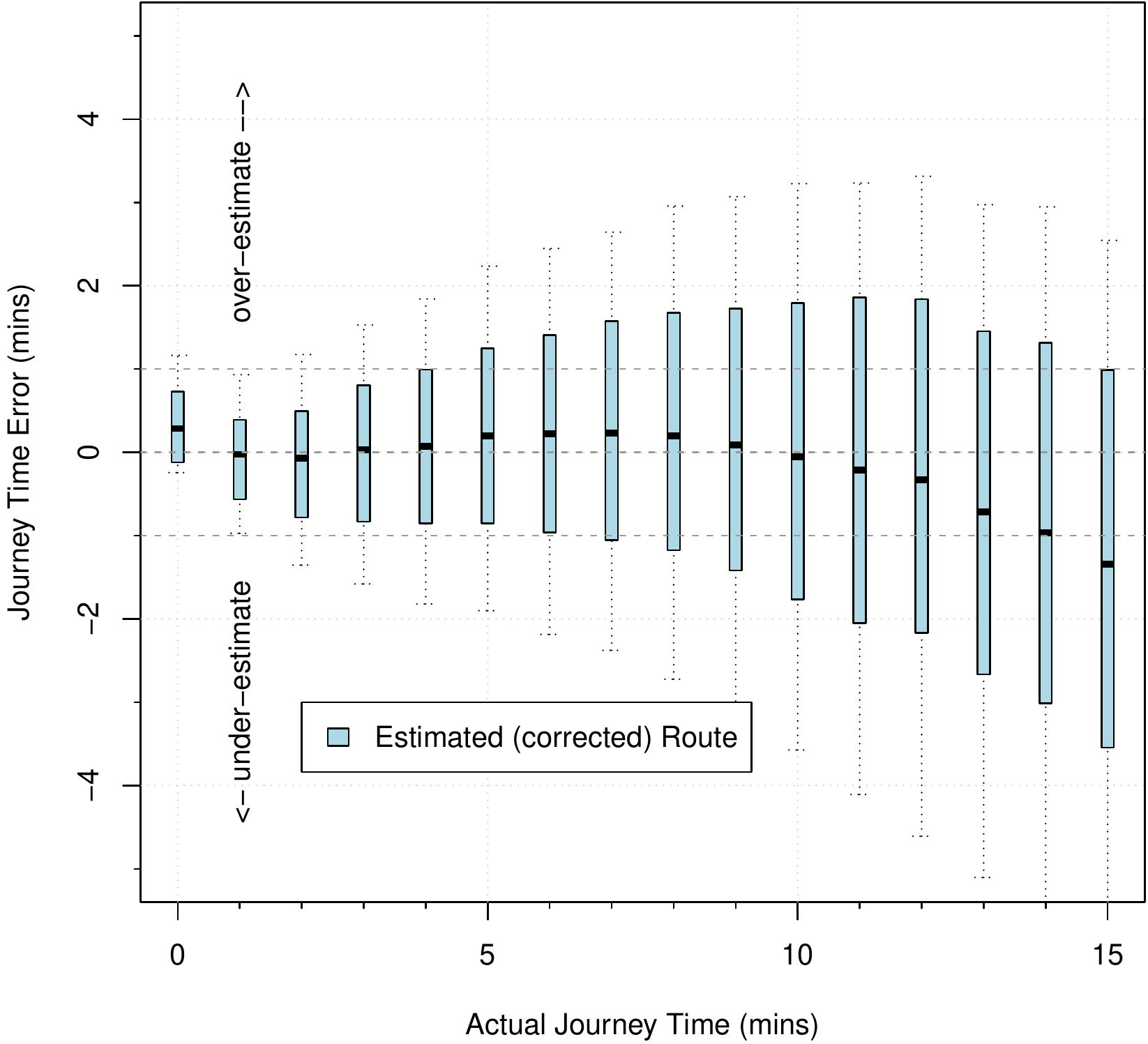}
{Arrival time prediction error using Metric V after correction. Boxes represent the 25th and 75th percentiles. Whiskers represent the 10th and 90th percentiles.  \label{fig:ra__e_rl_C.pdf}}

However, while the adjustment described above considerably reduces arrival time error, it does not improve route similarity scores. Recall from Figure \ref{fig:ra__simi-mean_U} that Metric II  achieves a similarity score of $80\%$ which compares favourably with $73\%$ obtained for Metric V. This suggests that a hybrid approach combining Metric II to select the route, subsequently employing Metric V to predict journey duration can yield increased accuracy both in arrival times and path similarity. Our objective in this case is to match the current performance of LAS rather than achieve the shortest possible arrival time.

To this end, first we optimise the static BLRN weights $W^{(\RN{2})}$ used by Metric II to maximise route similarity. The Nelder-Mead algorithm~\cite{nelder_simplex_1965} is employed on approximately $200$ routes selected from the LAS dataset with initial parameters set to the speeds employed by LAS (cf. Table \ref{tab:road_speed_las}) and allowing a maximum perturbation of $20\;\mathrm{mph}$ so a suitably wide range of alternative speeds can be explored. The resulting road speeds are displayed in Table \ref{tab:las_estimator_optimised}. Using Metric II with the Nelder-Mead optimised speeds we achieve a path coincidence of $84\%$, which represents best performance.


\begin{table}[!ht]
	\centering
	\caption{Road speed (in $\mathrm{mph}$ and junction delay (in $\mathrm{sec}$) for use with Metric II obtained by Nelder-Mead optimisation for route similarity.}
	\label{tab:las_estimator_optimised}
	\vskip .5cm
	\begin{tabular}{|l|r|r|}
		\hline 
		\textbf{Road Type} & \textbf{LAS } & \textbf{Nelder-Mead}\\ 
		\hline 
		Junction delay				& 2.5	&  4.33		 \\	\hline 
		Motorway					& 35		& 35.47		 \\	\hline 
		A Road					& 29		& 29.39		 \\ \hline 
		B Road					& 24		& 26.83		 \\	\hline 
		Minor Road				& 19		& 18.97		 \\	\hline 
		Local Street				& 14		& 15.51  	 \\	\hline 
		Alley						& 3		&  5.31		 \\	\hline 
		Pedestrianised Street		& 2		&  5.37		 \\	\hline 
		Private Road- Publicly 		& 5		&  8.37		 \\	\hline 
		Private Road- Restricted		& 5		&  6.84		 \\	\hline 
	\end{tabular} 
\end{table}

Following route selection using Metric II with the Nelder-Mead road speeds, we proceed calculate the estimated journey time using Metric V. Figure \ref{fig:ra__hybrid-1_.pdf} shows that this approach (green) outperforms Metric V after correction with the bias function (blue), achieving less than $60$ seconds accuracy for journeys up to $15$ minutes while maintaining best path coincidence of $84\%$. We refer to the method of route selection using Metric II with Nelder-Mead road speeds and subsequent journey time estimation using Metric V as the \textbf{hybrid model}, obtaining the best match to actual LAS performance.  

Using the hybrid model, the choropleth map in Figure \ref{fig:spatial-5_.pdf} depicts the spatial variation of the prediction error per London Borough. Figure \ref{fig:spatial-5_.pdf}  implies that the error is  lower in the suburbs, where there is a tendency to overestimate arrival times, while the reverse is true in the centre. 
The specific causes of these variations as well as the possibility of building area-specific mobility models will be investigated in future work. 

\Figure[!t]()[width=0.4\textwidth]{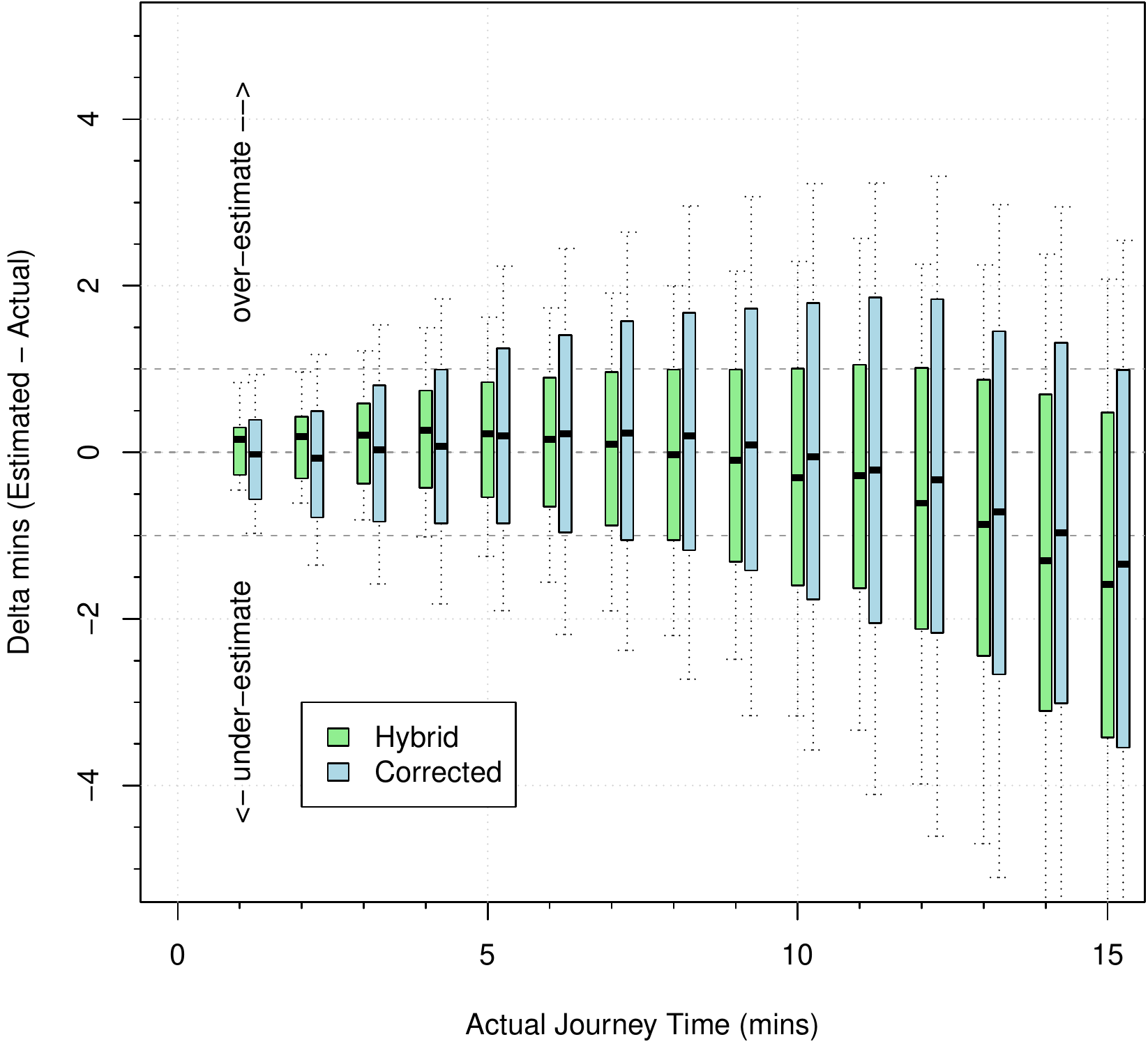}
{Arrival time error using the hybrid method.  Boxes represent the 25th and 75th percentiles. Whiskers represent the 10th and 90th percentiles. \label{fig:ra__hybrid-1_.pdf}}


\Figure[!t]()[width=0.4\textwidth]{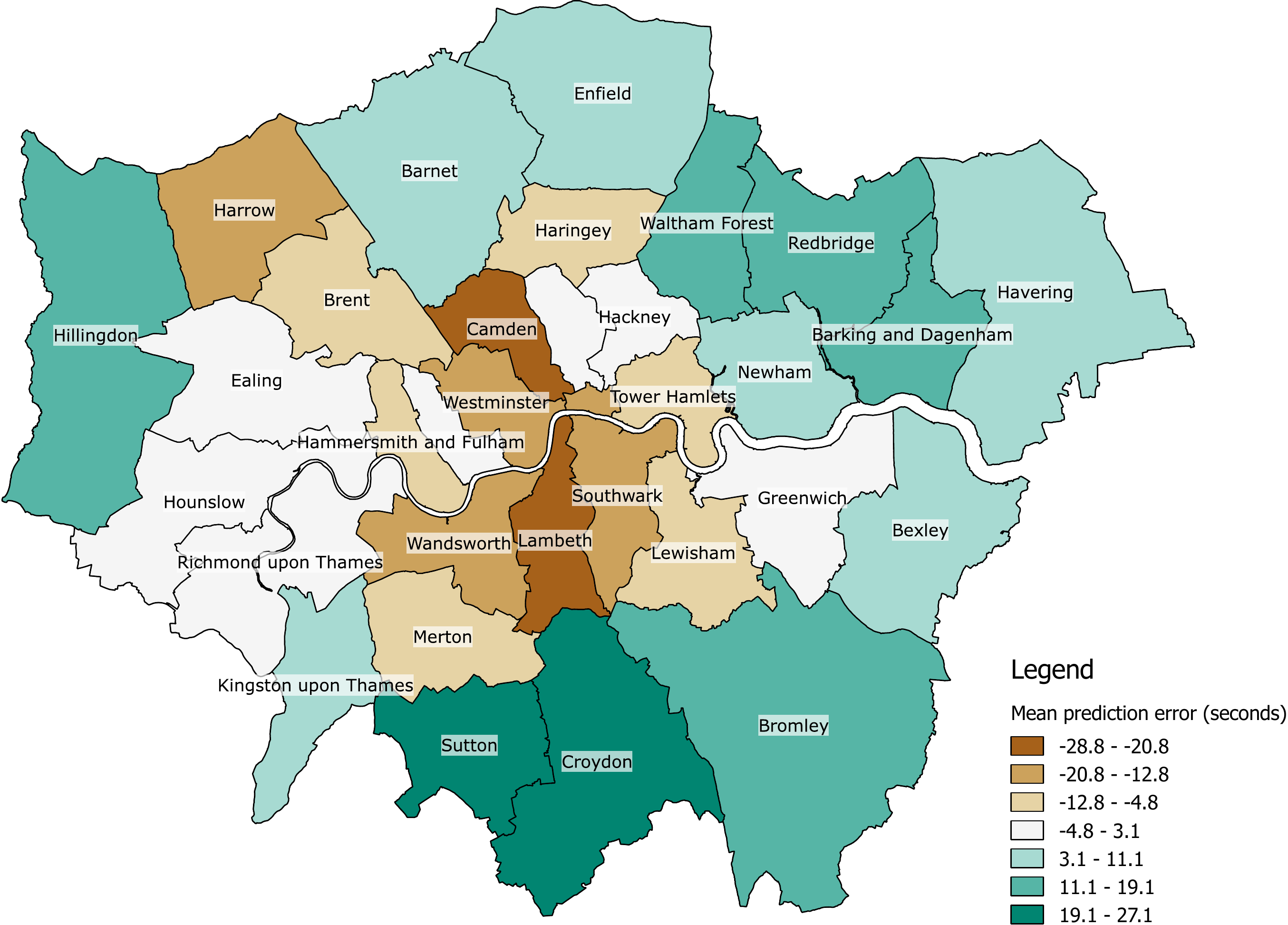}
{Mean prediction error using the hybrid method by London Borough. 	\label{fig:spatial-5_.pdf}}

\section{Discussion and Conclusions} 
This paper introduces a novel data-driven methodology for the accurate prediction of the route followed by an ambulance responding to an emergency incident travelling with blue lights and sirens on, and the precise estimation of its expected arrival time. Key ingredients of our approach include the comprehensive reconstruction of ambulance journeys from coarse location tracking data using map-matching; the development of a graph representation of the London road network specifically tailored to the particularities of emergency vehicles travelling with blue lights and sirens on; the estimation of several alternative edge-cost metrics for this network; the assessment of their performance characteristics; and, the use of the best performing metrics for the development of a hybrid model that achieves the highest route similarity while minimising arrival time error. This model offers consistent performance across both spatial and temporal dimensions.

This methodology has considerable implications for ambulance services aiming to improve the accuracy and fidelity of their emergency response emulations, which are the main instrument employed in practice for the investigation of more effective and more efficient operational policy. Specifically, the ability to trace closely the true movements of ambulances and to estimate their expected arrival time, enables the exploration of the advantages and limitations of alternative strategies  through emulation of realistic scenarios. In addition to forecasting, data-driven decision making can offer significant improvements in planning for key service requirements such as ambulance staffing levels and resource placement, and enables services to balance strategy and tactics through the accurate assessment, for example, of the effects of particular dispatch tactics on meeting their strategic objectives. Moreover, the methodology can be applied in real-time at emergency service operating centres for planning, to determine the ability of the service to cover its area of responsibility as well as to quantify the effects on its ability to maintain a high level of response resulting from specific vehicle dispatch or relocation decisions or as a result of permitting crews to go on a break.

One feature of this work is that our methodology considers only historical data collected internally by the emergency ambulance service. In future work, we aim to explore potential improvements that can be achieved using real-time information as well as traffic and related context information retrieved from external systems. Although extending the methodology to cater for such data sources appears relatively straightforward, a major challenge relates to assessing the quality of such third-party data sources and validating their accuracy on the fly, especially considering the effects on loss of life that erroneous or intentionally misleading information may have.

Finally, our analysis appears to suggest that a bespoke route and navigation engine tailored specifically to ambulances travelling with blue lights and sirens on, rather than one developed for general civilian traffic as is currently the case, can lead to significant reductions in crew arrival times at the site of an incident. Specifically, the use of Metric V introduced in this paper appears to select preferable ambulance routes that suggest a faster journey to the site of an incident. Clearly, extensive further work under true operational conditions is required to assess whether concrete performance benefits can be achieved following this approach.


\EOD

\end{document}